\documentclass[times, twoside, watermark]{zHenriquesLab-StyleBioRxiv}
\usepackage{blindtext}
\usepackage{color}
\usepackage{float}

\leadauthor{Morales} 

\begin{document}
\setlength\parindent{15pt}

\title{Representational Drift and Learning-Induced Stabilization in the Olfactory Cortex}
\shorttitle{Representational drift in the Olfactory Cortex}
%\title{Representational drift in the olfactory cortex and the role of learning}
%\shorttitle{Representational drift in the olfactory cortex and the role of learning}

% Use letters for affiliations, numbers to show equal authorship (if applicable) and to indicate the corresponding author
\author[1]{Guillermo B. Morales}
\author[1,\Letter]{Miguel A. Muñoz}
\author[2,\Letter]{Yuhai Tu}

\affil[1]{Departamento de Electromagnetismo y Física de la Materia and Instituto Carlos I de Física Teórica y Computacional, Universidad de Granada, E-18071 Granada, Spain}
\affil[2]{IBM T. J. Watson Research Center, Yorktown Heights, NY 10598}

\maketitle
%TC:break Abstract
%the command above serves to have a word count for the abstract
\begin{abstract}
%The mammalian brain, shaped by millions of years of evolution, encodes incoming stimuli through patterns of spiking activity, forming internal representations of the external world. While initially thought to be stable, recent studies challenge this assumption, revealing that neural representations can change over time in a phenomenon termed ``representational drift'' (RD). Here, we propose a biologically-realistic computational model of the mammalian olfactory system to elucidate the mechanisms underlying RD in the odor representation at the piriform cortex. Our model incorporates two mechanisms for the dynamics of synaptic weights operating at very different timescales: spontaneous multiplicative fluctuations on a scale of days and spike-time dependent plasticity (STDP) effects on a scale of seconds. Specifically, we show that multiplicative fluctuations in synaptic sizes induce drifting representations during baseline, spontaneous activity. On the other hand, STDP learning during stimulus presentation can explain why representations of already ``learned'' odors drift slower than unfamiliar ones, as well as the empirically observed dependence of the drift rate with the frequency of stimulus presentation. The proposed model not only offers a simple explanation for the emergence of drift and its relation to learning but also quantitatively reproduces several recent experimental results while suggesting new testable predictions. Altogether, our findings provide fresh theoretical insights into the dynamic nature of neural representations.

The brain encodes external stimuli through patterns of neural activity, forming internal representations of the world. Recent experiments show that neural representations for a given stimulus change over time. However, the mechanistic origin for the observed  ``representational drift'' (RD) remains unclear. Here, we propose a biologically-realistic computational model of the piriform cortex to study RD in the mammalian olfactory system by combining two mechanisms for the dynamics of synaptic weights at two separate timescales: spontaneous fluctuations on a scale of days and spike-time dependent plasticity (STDP) on a scale of seconds. Our study shows that, while spontaneous fluctuations in synaptic weights induce RD, STDP-based learning during repeated stimulus presentations can reduce it. Our model quantitatively explains recent experiments on RD in the olfactory system and offers a mechanistic explanation for the emergence of drift and its relation to learning, which may be useful to study RD in other brain regions.
\end{abstract}
%TC:break main
%the command above serves to have a word count for the abstract


\begin{corrauthor}
\texttt{mamunoz@onsager.ugr.es}
\texttt{yuhai@us.ibm.com}
\end{corrauthor}

\section*{Introduction}

The brain is a powerful computing machine, ``trained" by millions of years of evolution to process, represent, and interpret the thousands of incoming stimuli it is exposed to on a daily basis. The prevailing hypothesis suggests that the brain encodes information about such external inputs through patterns of neural spiking activity in sensory areas, often observed to reside within lower-dimensional manifolds \cite{chung_neural_2021,Stringer2019,Manley-1million}, which constitute an internal representation of the external world \cite{Abbott1999,Yuste2015}.

%Notably, when talking about RD, one needs to be careful and distinguish between two phenomenologically different types of RD, often sheltered under the same term. On the one hand,  there is experimental evidence of an input-evoked \emph{fast} drift, observed in a timescale of hours during representation learning of a new environment in the mouse hippocampus \citep{khatib_experience_2022}. This type of drift has been recently associated with an increase in the sparsification of the population code during learning \citep{ratzon_representational_2023}. On the other hand, there is evidence of a \emph{slow} drift that takes place over longer periods of weeks, even in the absence of stimuli \citep{schoonover_representational_2021}, for which the population statistics (average firing rate, sparsity of the responses, etc.) remain invariant. 

%When comparing the effects and properties of these ``drifting" neural codes across different brain regions, it is not uncommon to encounter experimental findings that appear contradictory, making it difficult to identify universal principles.
Nevertheless, when observed, the ubiquitous ``drift" of neural codes seems to differ in its properties across brain regions.  For instance, despite the measured changes at the single-cell level, overall population statistics have been shown to remain invariant across weeks in the piriform cortex \citep{schoonover_representational_2021} and posterior parietal cortex \citep{driscoll_dynamic_2017}, whereas in the hippocampus, drift at relatively short timescales of hours is associated with an increased sparsification of the population response \citep{khatib_experience_2022, ratzon_representational_2024}. Similarly, neural population responses to drifting gratings in mouse visual cortex have been shown to be stable across weeks, while encoding of natural movies in the same region appeared to change considerably across weeks, indicating the existence of a stimulus-dependent drift in visual cortex \citep{marks_stimulus-dependent_2021}. In contrast, drift rate in olfactory cortex was demonstrated to be fairly independent of the chemical nature of the odor but, remarkably, it could be slowed down by increasing the frequency of stimulus presentation \citep{schoonover_representational_2021}.

Representational drift (RD) is generally believed to be caused by changes at the synaptic level, which are difficult to measure, especially in behaving animals. Several recent studies have thus focused on studying RD by using computational approaches to model the dynamics of synaptic weights. In particular, several hypotheses have been investigated regarding the origin of RD, including: i) spike timing-dependent plasticity (STDP) or synaptic turnover, in combination with homeostatic normalization of synaptic weights \citep{kossio_drifting_2021}; ii) noisy synaptic-weight updates with white \citep{qin_coordinated_2023} or correlated \citep{rule_self-healing_2022} noise; iii) node or weight dropout \citep{aitken_geometry_2022}; and iv) implicit regularization of the population activity  \citep{ratzon_representational_2024}. Alternatively, it has been recently proposed that RD could be driven by fluctuations in the intrinsic excitability of neurons, rather than changes at the synaptic level \citep{delamare_drift_2023}. %However, although these theoretical studies provided plausible explanations for the emergence of RD using relatively simple models, \new{further evidence remains elusive due to lack of direct comparison with quantitative experimental data.}

In this paper, we focus on the mammalian olfactory cortex, for which new experimental findings on RD have been recently reported by Schoonover and colleagues using mice \cite{schoonover_representational_2021}. To draw a quantitative comparison with their experimental results, we develop a biologically realistic computational model of the mouse olfactory cortex, showing that a simple multiplicative stochastic process over the synaptic weights can account for the observed drift in the representation of odors, while naturally giving rise to the empirically measured log-normal distribution of weights and stable population statistics. Furthermore, our model can also explain why RD slows down when the frequency of stimulus presentation increases, as recently observed in experiments \cite{schoonover_representational_2021}.

To the best of our knowledge, our model provides the first quantitative explanation of the observed drift in the olfactory system. More importantly, an intuitive mechanistic picture emerges in which RD is caused by slow and spontaneous fluctuations in synaptic weights, while learning at a faster timescale drives the system deterministically towards a low-dimensional representation manifold, which effectively suppresses RD. We found that this general mechanistic picture,  although applied here in a relatively realistic model of the olfactory cortex, could also be useful in understanding RD in other regions of the brain.

\section*{The background: Odor encoding in the olfactory cortex}\label{Section_1_OdorEncoding}

To provide additional context for the problem, we first present an overview of the processes involved in odor perception in the mouse olfactory cortex, focusing on the neural architecture that underlies the internal representations of odors (see, for instance, \citep{blazing_odor_2020} for a recent review).

\begin{figure*}[!ht]
\centering{}
\includegraphics[width=0.9\linewidth]{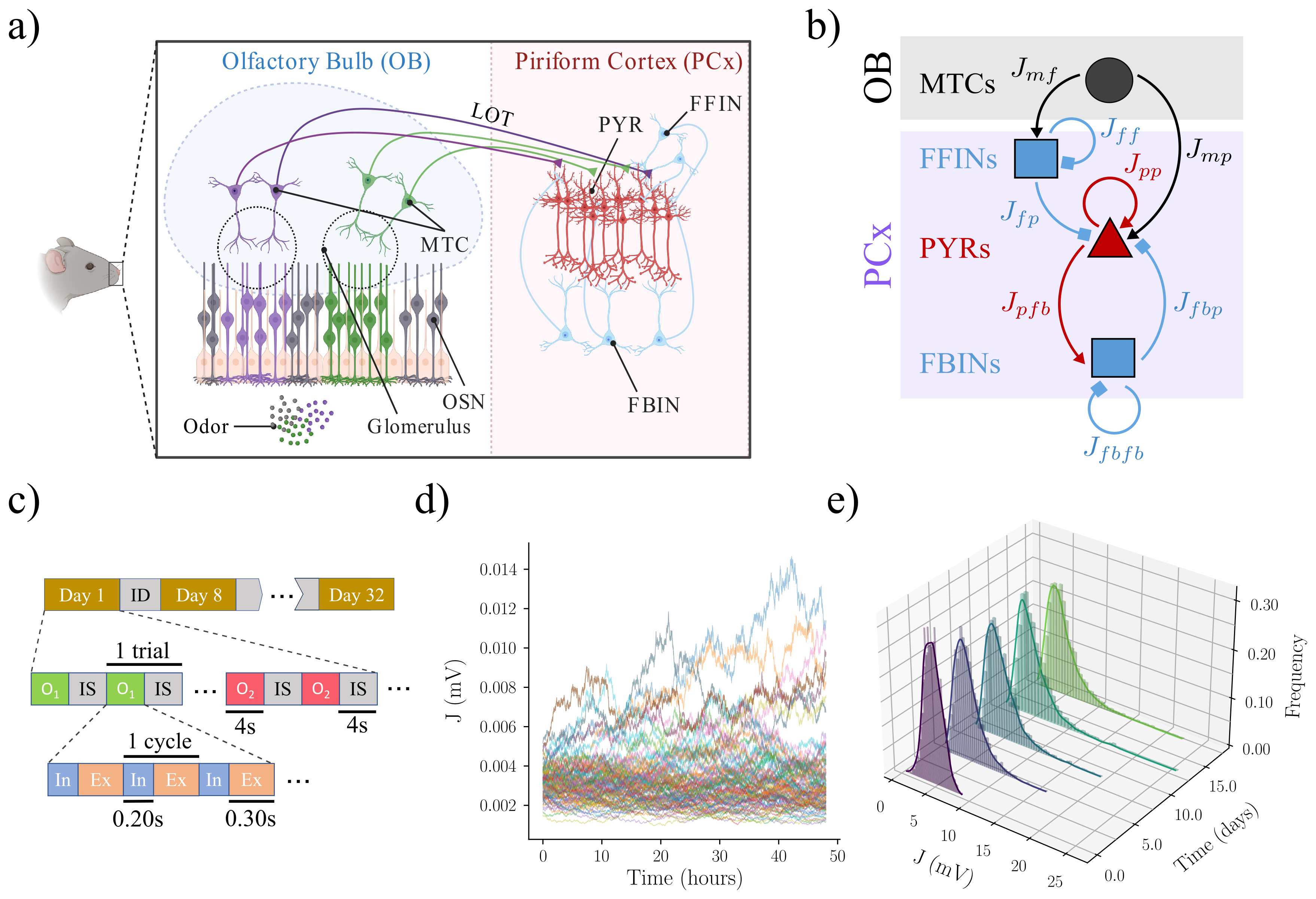}
\caption{\textbf{Model and experimental setup.} \textbf{(a)} When a given odor reaches the nasal epithelium, the different odorants composing it (green, purple and gray molecules in the figure) bind to specific receptors in olfactory sensory neurons (OSNs). Next, OSNs expressing the same type of odorant receptor project their axons into the same glomerulus, where they connect to the dendrites of mitral/tufted cells (MTCs). Random projections of MTCs axons into the piriform cortex (PCx) conform the lateral olfactory track (LOT). In the PCx pyramidal (PYR) neurons receive excitatory inputs from MTCs, as well as inhibitory connections from FFINs and FBINs. Illustration created using the BioRender software. \textbf{(b)} Diagram of the model, depicting the different types of neural populations considered and their interactions.  \textbf{(c)} Testing protocol for PCx responses to odor presentation, following the experimental setup in \cite{schoonover_representational_2021}. For each test day and trial, an odor $O_{i}$ is presented for $4\si{s}$, followed by an inter-stimuli (IS) transient of the same duration. During odor presentation, the spiking frequency of MTCs change during the inhalation (IN) and exhalation (EX) periods that constitute one respiration cycle (see main text). Test days are spaced in time by inter-days (ID) transients lasting $8$ days.\textbf{(d)} Trajectories of $100$ randomly chosen weights between MTCs and pyramidal neurons under the GMR process.  \textbf{(e)} Evolution of the probability distribution for the MTC-to-pyramidal weights, showing how an original Gaussian shape evolves towards a lognormal distribution. }\label{fig:1}
\end{figure*}

Odor perception begins when volatile molecules in the environment, known as odorants, bind to receptors in \emph{olfactory sensory neurons} (OSNs) of the nasal ephitelium during inhalation (see sketch in Fig.~\ref{fig:1}a). The pioneering work of Buck and Axell, back in 1991, revealed that each OSN in the mouse expressed just one out of $\sim 1000$ different odorant receptor genes  \citep{buck_novel_1991}.  Although these receptors exhibit high affinity for specific odorants, they are broadly tuned and can also bind to other volatile compounds with lower affinity. Nonetheless, olfactory sensory neurons (OSNs) with the highest affinity receptors for a particular odorant at a given concentration consistently fire first upon its presentation \citep{malnic_combinatorial_1999, jiang_molecular_2015}. 

During the next step of odor processing, all OSNs expressing the same type of receptor project their axons onto a unique set of glomeruli in the olfactory bulb (OB) \citep{mombaerts_visualizing_1996, halasz_terminal_1993} (Fig.~\ref{fig:1}a). These spheroidal structures host the synaptic connections between the OSN axon terminals and the dendrites of the secondary neurons: the \emph{mitral/tufted cells} (MTCs). For a given odorant, although OSNs with less specific receptors can be eventually activated, the MTCs associated with the most odorant-specific receptors will fire the earliest, effectively transforming the initial receptor-specific encoding into a temporal encoding in the OB. 

The information encoded in temporal patterns of MTCs activity in the OB is conveyed through random and overlapping \emph{lateral olfactory track} (LOT) connections into the piriform cortex (PCx), i.e., the \emph{temporal} encoding in the OB is further translated into an \emph{ensemble} code in the PCx, where odorant identity is determined by specific sets of principal neurons (pyramidal cells, mainly) recruited during the sniff %, with no other information about the temporal profile of the incoming spikes 
\citep{stern_transformation_2018}. Importantly, cortical odor responses are mostly determined by the earliest-active glomeruli (the ones with higher specificity for the given odorant) due to the fast recruiting of inhibitory interneurons (feed-forward inhibitory neurons (FFINs) and feed-back inhibitory neurons (FBINs), see Fig.~\ref{fig:1}a) that suppress the cortical response to later, less-specific OB inputs \citep{bolding_recurrent_2018, stern_transformation_2018, bolding_recurrent_2020,Meissner2023}. For simplicity, in the following results we will not  distinguish between "odors" and "odorants" and will use both terms indistinctly.

\section*{The modeling framework}\label{Section_2_ExpSetup}

\subsection*{A spiking network model of the olfactory cortex}\label{Section_2_Sub1_Modeling_PCx}

In order to make direct comparison with experiments, we developed a realistic model of the olfactory bulb (OB) and piriform cortex (PCx) based on the spiking network model proposed by Stern \emph{et al.} \cite{stern_transformation_2018}. As illustrated in Fig.~\ref{fig:1}c, the pyramidal neurons (PYRs) in PCx receive excitatory inputs directly from MTCs through LOT connections as well as recurrent connections with other pyramidal cells. The PYRs are also subject to inhibitory currents from FBIN and FFIN neurons, which receive excitatory inputs from PYRs and MTCs, respectively. 

RD occurs in a timescale of days, which is $\sim 10^4\times$ longer than the timescale studied in \cite{stern_transformation_2018} for fixed synaptic weights. To make computation for such long time scales feasible,  we reduced the number of neurons in our model but compensated it by increasing the connectivity so that, for each type of neuron, the average incoming excitatory and inhibitory currents remain the same as in \cite{stern_transformation_2018} so that the network remains balanced (see Methods and Fig.~\ref{fig:SI_DriftMeasures}j for details). The relative values of all network parameters (membrane timescales, average synaptic weights, fraction of responsive glomeruli, etc.) were kept as in \cite{stern_transformation_2018}. Following Stern \emph{et al.} \cite{stern_transformation_2018}, each respiration cycle consisted of an inhalation period of duration $\tau_{inh}=200\si{ms}$ followed by an exhalation period of duration $\tau_{exh}=300\si{ms}$ (see Fig.\ref{fig:1}b). MTCs respond with an enhanced firing rate to the presence of an odor only in the inhalation period. An example of the emergent pattern of MTC activity during the presentation of a particular odor is given in Fig.~\ref{fig:SI_DriftMeasures}a. To account for the baseline resting-state activity observed in the PCx in the absence of odor inputs, we included random Poissonian spiking of all pyramidal neurons at a relatively slow rate $f_{spont}=1 \si{Hz}$ (see Fig.~\ref{fig:SI_DriftMeasures}a). %We note that, when no stimulus is present, all MTCs have Poissonian spiking activity with a common baseline firing rate (see Methods).

\subsection*{Modeling dynamics of synaptic plasticity} \label{Section_2_Sub2_Modeling_Drift}
%Given the overall complexity of the odor encoding process, from the arrival of odor molecules in the nasal epithelium to the activation of pyramidal cells in the PCx, identifying the specific link in the chain from which RD emerges seems a daunting task. However, experimental findings can serve as guides and constraints for model building.

%Increasing evidence indicates that OB responses remain stable for long periods of at least several months \cite{bhalla_multiday_1997, kato_dynamic_2012, shani-narkiss_stability_2023}, implying that different odors are represented in stable temporal patterns of MTC activity. On the other hand, empirical \cite{bolding_recurrent_2020} and computational \cite{stern_transformation_2018} evidence shows that the identity of odors in PCx representations is fundamentally determined by the response of what we call \emph{``primary"} pyramidal neurons, that is, units that are directly recruited through excitatory OB inputs from early-responding glomeruli (in contrast, \emph{``secondary"} neurons can only be activated with the aid of recurrent excitatory interactions from other pyramidal neurons). Since Schoonover \emph{et al.} reported that only around $2.5\%$ of the pyramidal neurons recorded in PCx showed a stable response to the odor panel during the $32$ days of their experiments \cite{schoonover_representational_2021},  it must follow that \emph{any proposed RD mechanism in PCx must necessarily affect the identity of the primary pyramidal cells recruited by the MTCs}. 

The key new ingredient we introduced in our model is synaptic plasticity. Empirical evidence on the fundamental role of synaptic modifications in different parts of the olfactory cortex \citep{wilson_plasticity_2004,ito_olfactory_2008, ma_regulation_2012, cohen_differential_2015, jacobson_experience-dependent_2018,kumar_plasticity_2021} suggests that synaptic plasticity is the most plausible origin for RD. % rather than other potential mechanisms such as intrinsic plasticity (which involves changes in the excitability of individual neurons). 
However, in order to explain the observed RD behaviors in olfactory system, the responsible synaptic plasticity mechanisms need to satisfy several constraints: \textbf{(i)} give rise to stable log-normal-distributed values of the synaptic efficiencies, as it has been extensively documented by both \emph{in vivo} and \emph{in vitro} experiments \cite{loewenstein_multiplicative_2011,Buzsaki2014}; \textbf{(ii)} operate on a ``slow'' intrinsic time scale of days or weeks, in agreement with the observed time scale of the drift in the olfactory cortex \cite{schoonover_representational_2021}; \textbf{(iii)} lead to drifting representations of inputs across time; \textbf{(iv)} guarantee the empirically-observed invariance of population statistics despite the changing representations of the stimuli; and \textbf{(v)} explain the observed dependence of the drift rate with the frequency of stimulus presentation \cite{schoonover_representational_2021}. In our model, we incorporate a combination of two synaptic plasticity mechanisms that, together, can satisfy these constraints. 

\subsubsection*{I. The slow stochastic synaptic dynamics} It was shown by Loewenstein \emph{et al.} \cite{loewenstein_multiplicative_2011} that dendritic spines in the auditory cortex of mice exhibited substantial changes in size at timescales that ranged, precisely, from days to months. Moreover, not only could the stationary probability distribution of spine sizes be very well fitted by a log-normal distribution, but also the magnitude of change in spine sizes was found to be proportional to the size of the spines, hinting at the existence of an underlying multiplicative dynamics \cite{loewenstein_multiplicative_2011}. Inspired by these empirical findings, we propose to model the \emph{slow} dynamics of  synaptic weights as a \emph{geometric mean-reversion} (GMR) stochastic process \footnote{In order to preserve the sparsity in the connectivity matrices, only changes in already existing synapses were allowed.}:
\begin{equation}
    \dot{J}(t) = \omega (\mu - J(t))+ \sigma J(t) \xi(t),
    \label{eq:GMR_process}
\end{equation} where $J$ represents a non-zero synaptic weight between two neurons,  $\mu$ is the average value of the associated stationary log-normal weight distribution, $\xi(t)$ is a zero-mean, unit variance, Gaussian white noise, and $\omega$ and $\sigma$ are constants for the deterministic force and noise terms, respectively \footnote{It is shown in the SI that this synaptic plasticity mechanism can be approximated to a simplified version of the phenomenological one proposed by Loewenstein \emph{et al.} \cite{loewenstein_multiplicative_2011}.}. As an illustration, Fig.\ref{fig:1}d highlights the stochastic evolution of some LOT weights under our proposed rule,  whereas Fig.\ref{fig:1}e shows how multiplicative fluctuations evolve an initially Gaussian distribution for the weights towards a heavy-tailed (lognormal) stationary distribution.

\subsubsection*{II. The fast stimulus-dependent synaptic dynamics} From the Schoonover et al work \citep{schoonover_representational_2021}, the drift rate for representations of previously ``learned'' odors decreases with the frequency of stimulus presentation. This suggests the existence of a second mechanism concomitant to the previous one, operating on a much faster time scale of seconds (i.e., on the scale of the stimulus presentation). In this paper, we implemented a multiplicative STDP learning rule as proposed in \cite{rossum_stable_2000} (see Methods for details), which can alter synapses on a ``fast" time scale during odor presentation.

For simplicity, we only implement these two synaptic plasticity mechanisms in mitral-to-pyramidal (LOT) and pyramidal-to-pyramidal (recurrent) connections, while synaptic weights involving inhibitory neurons remain fixed. Furthermore, we neglect the effect of STDP during transient time without stimulus due to the fact that the background activity due to %nsWe also recall that, between experiments on different simulated days, PCx undergoes sustained resting-state activity that arises from both, 
random sparse inputs from baseline activity at the MTCs and low-rate random Poissionian spiking of pyramidal neurons (Fig.~\ref{fig:SI_DriftMeasures}a) is very small. % Since this pyramidal activity is markedly random, sparse, and weak , for computational convenience we disregarded the effect of STDP during these long transients between test days \footnote{We checked that STDP-mediated changes during transients are insufficient to explain the observed drift.}. 
For the same reason, given that the time spanned between test days is several orders of magnitude greater than the timescale of stimuli presentation, we also neglect the effect of the slow GMR process during the short periods of odor presentation.

\subsection*{Mimicking the experimental setup}\label{Section_2_Sub2_Modeling_Experiment}

In order to compare the results from computational analyses of our model with existing empirical evidence, we aimed at reproducing the experimental setup and analyses used by Schoonover \emph{et al.} \cite{schoonover_representational_2021}. 

Prior to the beginning of the experiment, we let MTC-to-pyramidal and pyramidal-to-pyramidal weights stochastically evolve under Eq.~\ref{eq:GMR_process} for $32$ days, to ensure that a quasi-stationary weight distribution was reached (see Fig.~\ref{fig:1}d and Fig.~\ref{fig:SI_DriftMeasures}b). Then, following the  experimental study, the total simulated experiment consisted of $32$ days during which the network dynamics is driven solely by the spontaneous pyramidal and MTCs activity. At $8$-day intervals, a test period is included during which $8$ different odors are sequentially presented.  More specifically, each odor is presented $7$ times (trials) during each test period, spanning $8$ respiration cycles ($4\si{s}$) in each trial, followed by a  $4\si{s}$ inter-odor transient between trials (see Fig.\ref{fig:1}b). 

For our analysis, we only consider neurons that show significant trial-averaged response over baseline activity to at least one odor in one day (Wilcoxon rank-sum test, $\alpha=0.005$). %On average, we find that $85.3 \pm  1.4\%$  of the pyramidal neurons meet this criteria (mean $\pm$ S.D.; $n=6$ realizations of the experiment). 
Specifically, for each test day $d$, odor $o$ and trial $m$, we averaged pyramidal responses in time across $2$ two-second windows after odor onset, then subtracted for each neuron its average baseline spontaneous rate to construct  population firing rates, $\mathbf{x}_{d,o,m}\in\mathbb{R}^{2N_{PYR}}$. Unless otherwise stated, all measures are performed on such spontaneous, baseline-subtracted population vectors of pyramidal activity, which we refer to as \emph{representations}. 

Finally, to identify the primary pyramidal neurons (i.e., those directly activated by the OB, without the need for recurrent excitation), we simulated again the first and last test days under the exact same conditions, but setting to zero all pyramidal-to-pyramidal synaptic efficacies. Secondary neurons for a given odor and day were then identified as those that showed over-baseline spiking activity in the original experiment, but were unresponsive after removing intracortical connections. 

\begin{comment}
  \begin{figure}
\centering{}
\includegraphics[width=0.9\linewidth]{Figures/Figure_2_Final.png}
\caption{\textbf{Sensitivity of individual neurons to odors changes over $32$ days.} Raster plots (top) and trial-averaged instantaneous firing rates (bottom) for three pyramidal neurons in response to a given specific odor, measured on each of the test days. Each of the seven rows in the raster plot represents a single trial of odor presentation. The dotted line indicates the end of stimulus presentations after 4 seconds. }\label{fig:PSTH_Curves}
\end{figure}  
\end{comment}

\begin{figure*}
\centering{}\includegraphics[width=15cm]{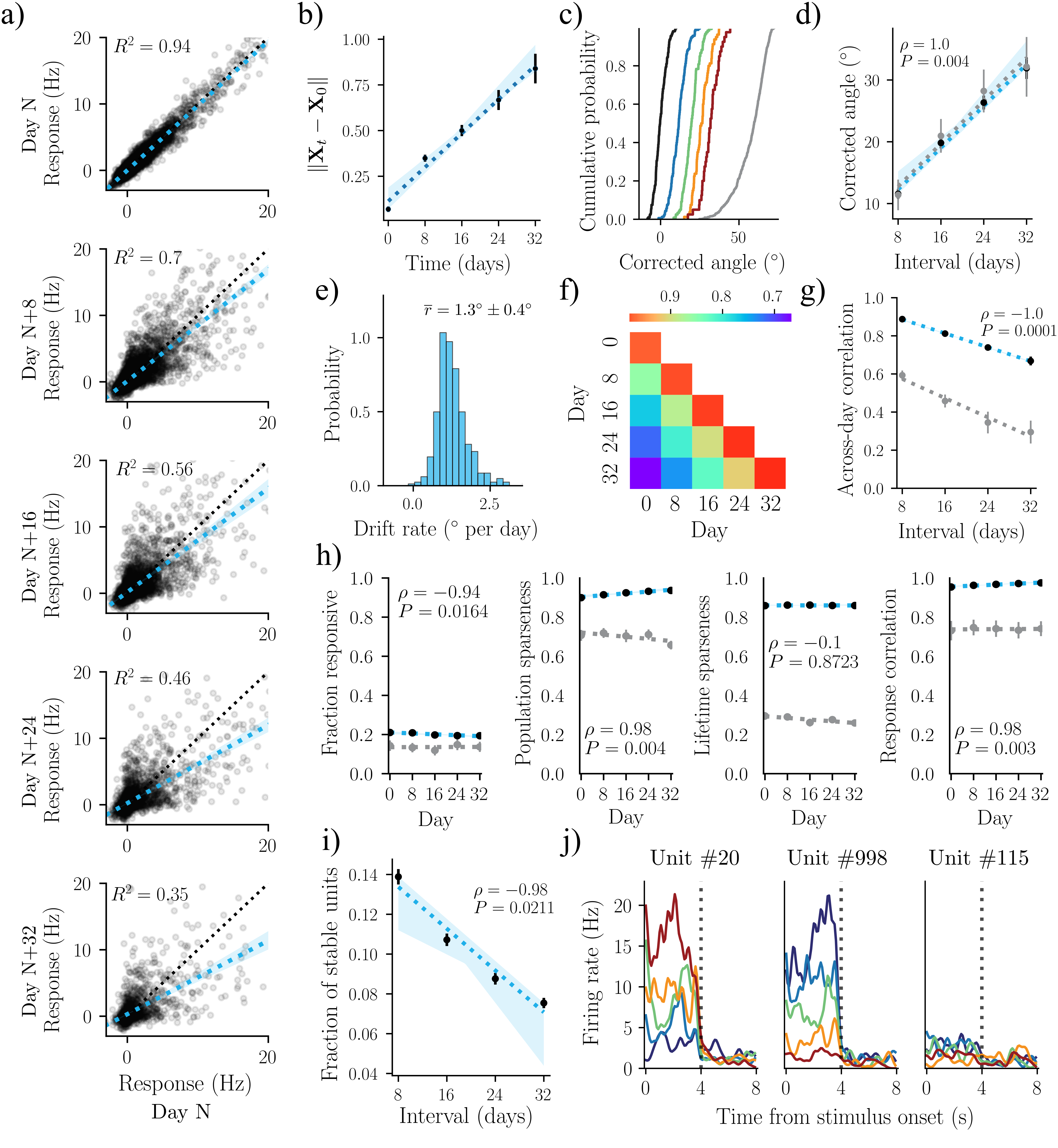}
\caption{\textbf{Odor representations drift despite invariant population statistics.} \textbf{(a)} Regression of firing rate responses within day (even vs odd trials, top panel) and across $8$- to $32$-days intervals for 500 randomly chosen odor-unit pairs. Black dashed line indicates identical response. \textbf{(b)} Euclidean distance between same-odor representations on first and later test days, averaged over odors and trials, and normalized by average within-day distance for different odor representations.\textbf{(c)} Cumulative probability distribution for the corrected angle between same-odor representations on a given day (black), across $8$-, $16$-, $24$- and $32$-day intervals (blue, green, yellow and red, respectively), and between different odors on the same day (gray). \textbf{(d)} Evolution of average corrected angle with interval between test days.  \textbf{(e)} Histogram for the observed drift rate in degree angles per day.  \textbf{(f)} Average population vector correlations for same-odor representations across days.  \textbf{(g)} Average correlation decay against time interval between test days. \textbf{(h)} Population statistics, including, on each test day and from left to right: fraction of responsive neurons (Wilcoxon rank-sum test, $\alpha=0.005$); average population sparseness; average lifetime sparseness and average within-day correlations. \textbf{(i)} Fraction of pyramidal neurons that show a stable response to a given odor across the experiment, averaged over all odors. \textbf{(j)} Trial-averaged instantaneous firing rates for three pyramidal neurons in response to a given odor, measured on days $0$ (purple), $8$ (blue), $16$ (green), $24$ (yellow) and $32$ (red) . Dotted line marks the end of stimulus presentation.  For all plots, when measures are reproduced from \cite{schoonover_representational_2021}, experimental results (gray markers) are compared with our simulations (black markers). Error bars were computed as the standard deviation across $n=6$ realizations of the experiment (i.e., across mice in the experimental results and different initial conditions in the simulations). Shaded blue regions in linear regressions represent $95\%$ confidence interval.}\label{fig:2}
\end{figure*}

\section*{Results: Quantitative comparison between model and experiments} \label{Section_3}

\subsection*{Representational drift in the PCx} 

%\section*{Representational drift in the PCx caused by the slow stochastic synaptic dynamics.} \label{Section_3}

Using our model, we studied characteristics of RD systematically and compared our results with experimental measurements by Schoonover et al \cite{schoonover_representational_2021}. First, we focused on the single-unit firing rate responses for each odor across days. As shown in Fig.~\ref{fig:2}a, we found the responses to the same odor became increasingly dissimilar over time. The squared Pearson-correlation ($R^2$) decreases with time interval: $R^2 = 0.94, 0.70, 0.56, 0.46, 0.35$ for within-day, 8-day, 16-day, 24-day, 32-day intervals, respectively, which quantitatively agrees with experimental results~\cite{schoonover_representational_2021}. %As expected, these changes in single-unit responses caused odor representations to drift in the high-dimensional space of all neurons activity, as 
The drift can also be measured by the average distance between a given odor representation on the first test day and the representation of the same odor on a later day as shown in Fig.~\ref{fig:2}b. 

%Now, given that we followed in our simulations the exact experimental protocol in \cite{schoonover_representational_2021},  computing also the same metrics gave us the possibility to make a sensible comparison with the empirical values, starting with 
Next, we analyzed the angle between trial-averaged population vectors (corrected for within-day variability; see Methods) on different test days in our model and compared it to the experimental measurements. In Fig.~\ref{fig:2}c, we showed the accumulative distributions of the normalized angles between representations for the same odor but on different test days as well as the angle between representations of two different odors measured on the same day (gray line) for reference. It is clear from Fig.~\ref{fig:2}c that as the time interval increases, the cumulative distributions shift to the right indicating an increase in the angle. However, even for the longest time interval of 32-days (red line in Fig.~\ref{fig:2}c), the same-odor angle is still smaller than the same-day different-odor angle (gray line). Quantitatively, the average angles for different interval times are in excellent agreement with experimental measurements~\cite{schoonover_representational_2021} as shown in Fig.~\ref{fig:2}d (black symbols: simulations; gray symbols: experiments). Moreover, measures of the average drift rate per day (see Eq.(\ref{eq:Ch4_DriftRate}) in Methods) resulted in a distribution with  $\overline{r}=1.3^{\circ} \pm 0.4^{\circ}$ (Fig.~\ref{fig:2}e), which is in quantitative agreement with the experimental result~\cite{schoonover_representational_2021}:  $\overline{r}_{exp}=1.3^{\circ} \pm 1.2^{\circ}$. Pearson correlations between all trial-averaged population responses to a common odor were also computed across all possible pairs of test days (Fig.~\ref{fig:2}f), showing that the average pair-wise correlation decreased with the time span between representations (Fig.~\ref{fig:2}g) in agreement with experiments. 

To rule out the possibility that the observed changes in odor-evoked responses originate from changes in overall population activity (for instance, the network becoming more reactive to stimuli), we studied the population-level statistics of the system. %Unlike other proposed models for drifting representations that led to sparsification of the population responses \cite{ratzon_representational_2024}, or did not respect the sparsity of neural connectivity \cite{kossio_drifting_2021}, 
We found that despite the drifting response of individual neurons, the population-level statistics (see Methods section) remain stable across days. As shown in different panels in Fig. \ref{fig:2}h, the fraction of responsive neurons (see also Fig.~\ref{fig:SI_DriftMeasures}c), population sparseness (Eq.~(\ref{eq:SI_Ch4_PopSparseness})), lifetime sparseness (Eq.~(\ref{eq:SI_Ch4_LTSparseness})), and within-day correlations (Eq.~(\ref{eq:SI_Ch4_WithinDayCorr})) remain stable over the 32-day period, consistent with experimental observations~\cite{schoonover_representational_2021}. 

%\new{Here, I would keep some of the original explanation, to give a sense of better understanding regarding why we do not reproduce the exact experimental values.}

The quantitative discrepancies with the experimental values may be caused by the particular choices of model parameters such as the size of the system, the sparsity of LOT connections, and the way we modeled the odors at the OB level. For example, in our model, odor responses involve on average a smaller fraction of the total PCx population (greater population sparseness), and the responsive neurons are specifically tuned to a small numbers of stimuli (greater lifetime sparseness; see Fig.~\ref{fig:SI_DriftMeasures}d in the SI).   However, these quantitative differences do not affect the general conclusion regarding stability of the population-level statistics. %  resulting also in a broader tuning for the pyramidal cells). %In any case, the above results suggest that the general properties of odor encoding changed only marginally over time, as observed in the experiments (Fig. \ref{fig:2}h, gray markers). 
\begin{figure*}
\centering{}
\includegraphics[width=\linewidth]{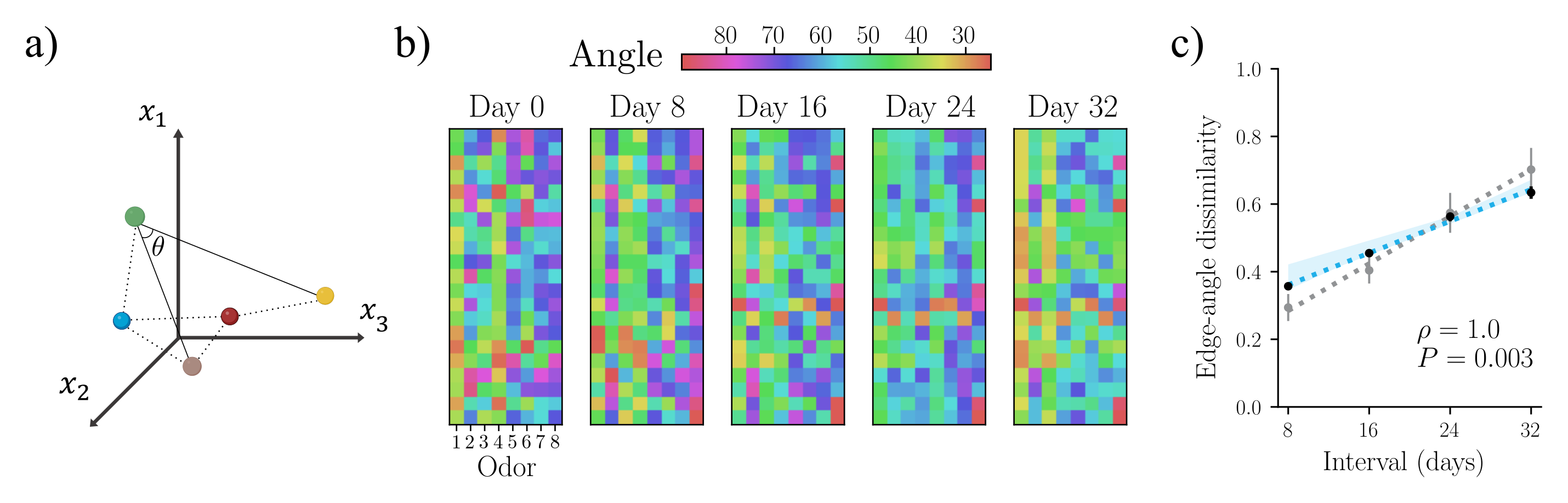}
\caption{\textbf{The geometry of odor representation manifolds changes with time.} \textbf{(a)} Schematic depiction of a hypothetical representation manifold geometry, marking one of the angles defined by the population vector responses to three different odors. \textbf{(b)} Edge-angle similarity matrices at test days. Note that for each odor/node there are $21$ possible associated angles, which define the number of rows in the similarity matrix
\textbf{(c)} Edge-angle corrected dissimilarity measure as a function of the time interval between test days ($\rho$ and $P$  denote the correlation coefficient and $p$-value for the linear regression). Error bars were computed as the standard deviation across $n=6$ realizations of the experiment (i.e., across mice in the experimental results and different initial conditions in the simulations).}\label{fig:4}
\end{figure*}

%The highly dynamic nature of neural representations suggests that individual neurons do not remain responsive to a given odor permanently. Indeed, as shown in Fig.~\ref{fig:2}i, the fraction of neurons showing a stable response to a given odor decreases significantly over a timescale of days. For example, neuron \#998 in Fig.~\ref{fig:2}j has lost its response to an odor over the 32-day interval. However, given the statistical stability of the neural representation, some neurons gain sensitivity to the odor, e.g., neuron \#20 shown in Fig.~\ref{fig:2}j. Of course, given the sparse representation of the odor, some neurons remain non-responsive to the odor, e.g., neuron \#115 in Fig.~\ref{fig:2}j.  
%\new{If we remove all this we are missing a lot of results in the SI that never get mentioned anywhere in the text... I rewrote the original paragraph to make it lighter while keeping all references too figures  (see commented section below)}
%We found that, as it was empirically reported by Schoonover and colleagues \cite{schoonover_representational_2021}, 
As expected, drift in cortical representations was manifested at the single-neuron level as observed in the experiments~\cite{schoonover_representational_2021}. In particular, pyramidal neurons fall into several categories in terms of their responses to the odors over time: (i) neurons that gained sensitivity to an odor (Fig.~\ref{fig:2}h and Fig.~\ref{fig:SI_DriftMeasures}f, unit \#20; Fig.~\ref{fig:SI_DriftMeasures}g, units highlighted in blue); (ii) initially responsive neurons that eventually lost their responsiveness (Fig.~\ref{fig:2}h and Fig.~\ref{fig:SI_DriftMeasures}f, unit \#998; Fig.~\ref{fig:SI_DriftMeasures}g, units highlighted in red); and (iii) neurons that showed a relatively stable response across all days (Fig.~\ref{fig:2}h and Fig.~\ref{fig:SI_DriftMeasures}f, unit \#115). Quantitatively, only $1.92\pm0.35\%$ of all considered pyramidal units showed a stable response (Wilcoxon rank-sum test, $\alpha=0.005$) across the full panel of odors (experimental value: $2.5\pm0.5\%$ \cite{schoonover_representational_2021}). Similarly, the percentage of pyramidal units maintaining a stable response to a given odor was $7.3 \pm 0.9 \%$ (Fig.~\ref{fig:2}i), also in perfect agreement with the experimentally measured value ($6.6 \pm 0.9 \%$ \cite{schoonover_representational_2021}). Thus, in line with the experiments, our model induces progressive changes in the selectivity of units which accumulate over time across the duration of the experiment (Fig.~\ref{fig:SI_DriftMeasures}h).

From our model, we can also identify two types of pyramidal neurons responsive to a given odor: primary neurons, which are directly excited by the OB, and secondary neurons, which are recruited by the primary neurons through the recurrent connections. While the relative fraction of primary and secondary responsive units is stable across days (Fig.~\ref{fig:SI_DriftMeasures}e), we found that primary neurons are relatively more stable across the 32-day experiment than secondary ones (Fig.~\ref{fig:SI_DriftMeasures}i). 

\subsection*{Geometry of the representation space}

%Although beyond the scope of this work, the question still remains of how a consistent perception of an objective reality (i.e., a given odor in this case) could emerge from the readout of constantly changing internal representations in sensory regions. One possibility, for instance, considers that downstream regions performing such a readout could adapt to the drift provided there was some invariant geometry of the representation manifold (for example, if all induced changes could be mapped to a rotation or other type of transformation of an invariant manifold in a high-dimensional space \cite{qin_coordinated_2023}).

%It has been suggested in previous studies that
Despite the observed drift for individual odors, it has been suggested that the relative positions of different stimuli in the representational space may remain invariant~\cite{qin_coordinated_2023}. To test whether such an invariant geometry was present in our model,  we computed the relative angles among odor representations following the same methodology as in \cite{schoonover_representational_2021}. %More specifically, if one considers the mean response of all recorded neurons over the duration of the odor presentation as a point in an $N$-dimensional space, then all points corresponding to different odors would span a certain response manifold 
More specifically, for each individual day $p$, an odor similarity matrix $A^p$ can be computed. As shown in Fig.~\ref{fig:4}b, each column corresponds to an individual odor (e.g., the green dot in Fig.~\ref{fig:4}a) and each row corresponds to a pair of other odors (e.g., the yellow and gray dots in Fig.~\ref{fig:4}a). Thus, each matrix element of $A^p$ is defined as the cosine similarity angle $\theta$ spanned by three odors in the representation space (see Methods for details).

As shown in Fig.~\ref{fig:4}b, the similarity matrices for different days change significantly with time, indicating the absence of a time-invariant geometrical structure in the odor representation space. As a way of quantifying this change in the geometry (relative position) of the representation space for different odors, we measured the \emph{matrix dissimilarity} ($\lVert A^{p,q}  \rVert_{F} \coloneqq \lVert A^p - A^q  \rVert_{F}$, where $F$ stands for Frobenius norm) between any two test days $p$ and $q$ (see Methods). Fig.~\ref{fig:4}c shows the value of this quantity against the time interval between the considered similarity matrices. In agreement with experimental results (gray markers), changes in edge angles between encoded odor responses accumulate over time, indicating a lack of geometrical invariance in the representation space for the olfactory system.

%\new{I would go for the old version in the following paragraph as well, as we are not really covering this aspect in the Discussion either, and I feel it makes more sense to provide thetechnical details" here}

However, although there is a lack of strict geometrical invariance with respect to the relative odor angles, the odors are always separable in the representational space. Furthermore, as we will describe in the Discussion section later in this paper, some geometrical properties of the odor representations such as the dimensionality of representations are found to be statistically invariant over time. 

% Despite the above conclusions, it is also remarkable that, regardless of the obvious changes in pair-wise correlations (Fig.~\ref{fig:2}f-g), it was always possible to cluster together the most correlated units in each test day to find a clear pattern in the correlation matrix that highlighted each odor representation (Fig.~\ref{fig:SI_DriftMeasures}k, hierarchical clustering with complete linkage; see Methods). This suggests that, for a given set of stimuli, even when the representation manifold presents an inconsistent geometry across days, some of the population encoding properties remain invariant. As a preliminary result in this direction, we found dimensionality (measured as the participation ratio using the response covariance matrix eigenspectra, see Methods) to be one of such properties (Fig.~\ref{fig:SI_DriftMeasures}l). 

\subsection*{The effects of stimulus-dependent plasticity}\label{Section_4}

\begin{figure}[!ht]
\centering{}
\includegraphics[width=\linewidth]{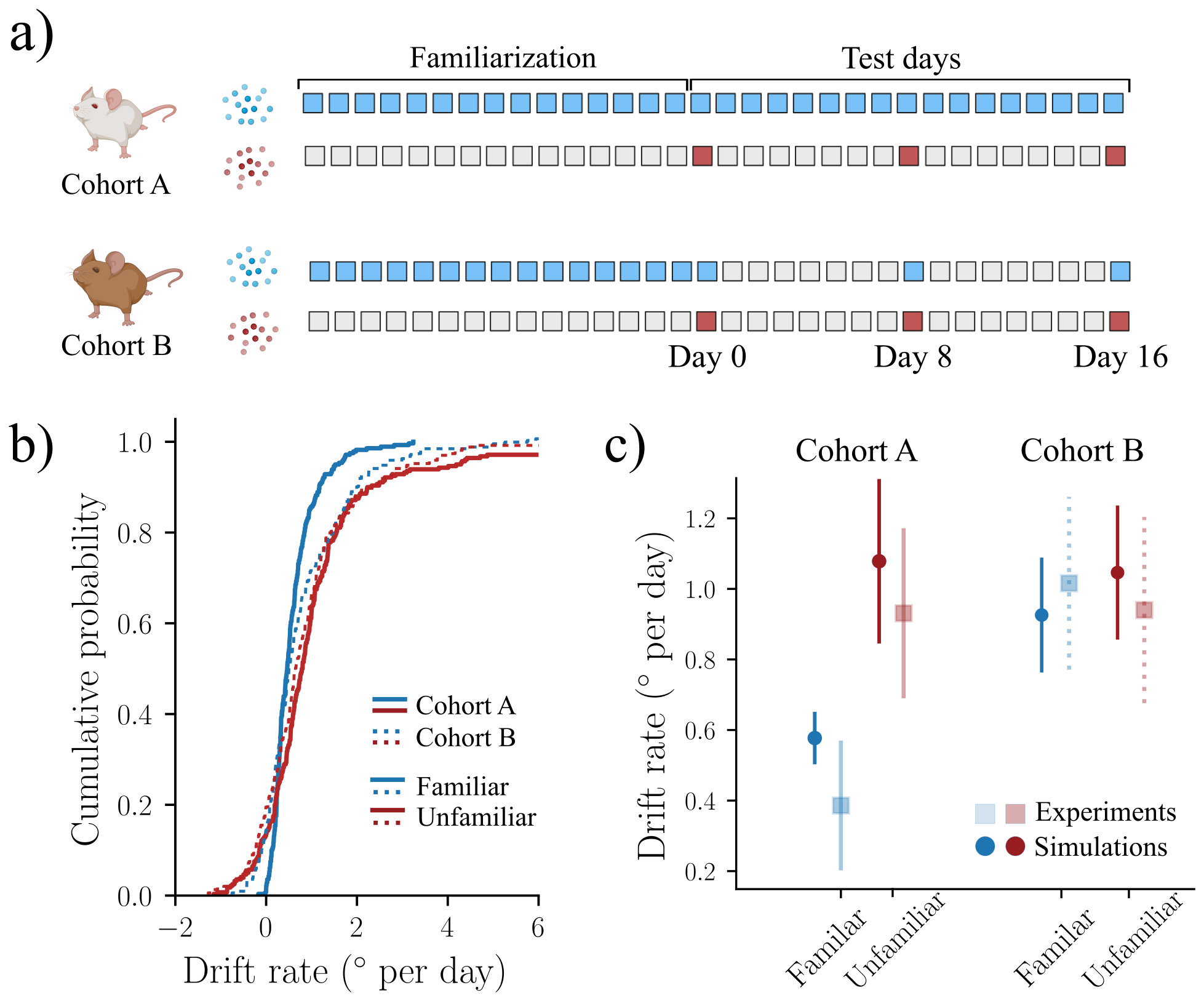}
\caption{\textbf{Representational drift depends on the frequency of stimulus presentation.} \textbf{(a)} Experimental setup to assess the dependence of the drift on the frequency of stimulus presentation, showing the two different cohorts of mice considered in the original experiment.  \textbf{(b)} Cumulative probability distribution for the drift rate in each simulated ``cohort'' and across familiar and unfamiliar odors. \textbf{(c)} Mean drift rate (in degrees per day) across odors in each of the simulated experiments, with the experimental values from \cite{schoonover_representational_2021} plotted for comparison. For each condition, values were averaged over 8 odors and $n=5$ ``mice" (i.e., different initial networks), with error bars representing a 95\% CI. }\label{fig:5}
\end{figure}

So far, we only considered the effects of the slow stochastic synaptic dynamics for RD. We now turn our attention to the effects of learning on RD motivated by the insightful experiments by Schoonover et al. \cite{schoonover_representational_2021}. %As we have seen, the empirically observed effects of RD over the encoding of odors in the PCx can be successfully explained by assuming  a GMR stochastic process in the weight dynamics. While for such analyses the presented odors had not been shown to the mice before the first test day, a yet more intriguing phenomenon was observed empirically when measuring the drift in the representations to odors that were familiar to the mice.
Specifically, in \cite{schoonover_representational_2021} a cohort A of $n=5$ mice were presented with a panel of $4$ odors daily across $16$ days prior to the beginning of the experiment. Beginning on day $0$, the same set of already ``familiar'' odors was still presented on a daily basis (Fig.~\ref{fig:5}a, cohort A, blue odors), but mice were also subject to a set of  four ``unfamiliar'' odors at 8-day intervals (Fig.~\ref{fig:5}a, cohort A, red odors). Interestingly, a slower drift rate for the representations of familiar odors (i.e., those presented daily on the 16 days prior to the experiment) was observed when compared to the drift for the unfamiliar ones. Notably, a second cohort of mice in which the familiar odors were not presented daily after day 0, but at 8-day intervals instead (Fig.~\ref{fig:5}a, cohort B, blue odors), showed no statistically significant changes in the drift rate with respect to the unfamiliar ones. Taken together, these results suggest that ``learned" representations of familiar stimuli would naturally drift as rapidly as representations of new inputs unless the familiar stimuli are presented with a higher frequency.

Here, we test whether our model, which incorporates the fast stimulus-dependent plasticity governed by the STDP learning rule, can explain these experimental observations by following the same protocols used in the experiments (Fig.~\ref{fig:5}a~\footnote{In order to track the weight dynamics, we reduced the network size and adjusted the density and average weight of connections accordingly to keep our simulations computationally feasible (see Methods)}) % which agree with experiments quantitatively.  %shows the model results for the familiar and unfamiliar stimuli averaged over the same set of 8 different odors mimicking the conditions for both cohort A and B. 
For our simulations of cohort A, we saw that angles between same-odor representations measured on different days are larger in the unfamiliar odor case than in the familiar odor case (Fig.~\ref{fig:5}b, see also Fig.~\ref{fig:SI_FamVSUnf}d in SI). Quantitatively, this translates into an average drift rate for the unfamiliar odor that is twice as fast ($\overline{r}_{unf}=1.1  (0.9-1.3)^{\circ}~ \mathrm{per~day}$), as compared to the case when in which familiar odors were presented daily ($\overline{r}_{fam}=0.58  (0.64-0.53) ^{\circ}~ \mathrm{per~day}$),  in excellent agreement with the experiments as shown in Fig.~\ref{fig:5}c. 
In the case of familiar odors presented every test day, we found that responses of individual neurons were more stable (Fig.~\ref{fig:SI_FamVSUnf}b),  showing slowly decaying correlations (Fig.~\ref{fig:SI_FamVSUnf}c) and smaller average distances between across-days representations (Fig.~\ref{fig:SI_FamVSUnf}e). 

We conducted additional simulations changing the interval between stimulus presentations from $1$ to $8$ days (thus interpolating between the conditions for familiar odors in cohort A and B). Our model results show that RD continuously decreased with the stimulus presentation frequency. 
While the reduction is minimal for the 8-day presentation interval, it increases to $\sim 30-40\%$ for the 1-day presentation frequency (see Fig.\ref{fig:5}c and Fig.~\ref{fig:SI_FamVSUnf}f in SI). We also verified that the model results were not due to the reduced system size, nor the particular new set of parameters (see Fig.~\ref{fig:SI_FamVSUnf}g in SI). 

%have a gradually decreasing impact on the overall drift rate as the gap between test days increases (Fig.~\ref{fig:SI_FamVSUnf}f). \new{the use of "plasticity" is a bit imprecise here, I think.}

%To understand this phenomenon, let us now go back to the representation manifold picture and consider that, through continued exposure to a particular odor, synaptic plasticity mechanisms at both, pyramidal-to-pyramidal and LOT synapses lead to a learnt representation, $\mathcal{D}_0$, of the given odor in the PCx activity, characterized by certain stationary probability distributions of the weights. As shown before, during the days in which the odor is not being presented, noise-induced fluctuations in the strength of synapses can cause the original representation, $\mathcal{D}_0$, to drift in the activity space towards a new representation, $\mathcal{D}_T$, after a certain time $T$ has elapsed (Fig.~\ref{fig:6}a). There are fundamentally two contributions to these changes: (i) a deterministic ``force'' (first term Eq.(\ref{eq:GMR_process})) that pushes large weights towards the average value of the distribution; and (ii) a random diffusion (second term Eq.(\ref{eq:GMR_process})) that induces multiplicative fluctuations in the weights. 

\begin{figure}[!ht]
\centering{}
\includegraphics[width=\linewidth]{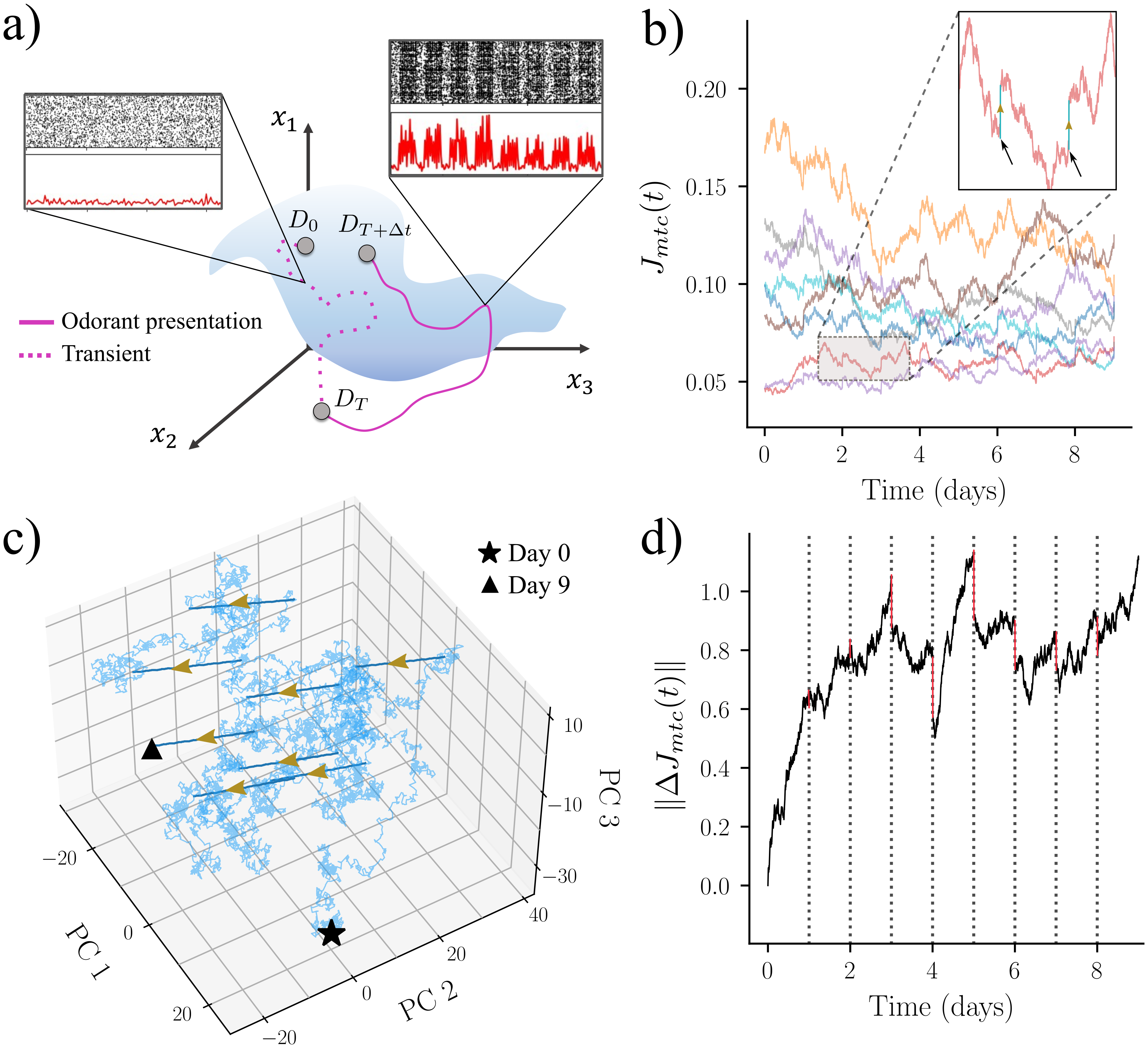}
\caption{\textbf{A mechanistic picture of learning-dependent drift.} \textbf{(a)} Schematic depiction for the evolution of an initial odor representation, $D_0$, to a new representation, $D_{T+\Delta t}$, reached after a long transient of length $T$ and the presentation of the stimulus for a time span $\Delta t \ll T$ . \textbf{(b)} Evolution of some weights during the simulated experiment with a familiar odor presented daily. Inset: close-up of a particular weight evolution, with black arrows pointing to the times of odor presentation. The large changes in weights during the odor presentation are highlighted by green arrowed lines. Weights were sampled every $\Delta t = 60 \si{s}$ during transient, and $\Delta t = 50 \si{ms}$ during stimulus presentation. \textbf{(c)} Projection of the weights dynamics into the first three principal components resulting from applying PCA during the presentation of the odor on the first day. Slow random fluctuations during inter-tests transients (light blue) are followed by fast, directed changes  during odor presentation (dark blue). Arrows indicate time direction during the experiment. \textbf{(d)} Normalized distance, in the projection space spanned by the first 10 principal components, between the initial weight configuration and the weights at time $t$. Dotted vertical lines mark the time of odor presentation, coinciding with sharp drops in the measured distance (highlighted in red).}\label{fig:6}
\end{figure}

Verified by its agreement with experiments, our model can be used to gain insights about the underlying mechanism for how learning suppresses RD. By systematically probing the dynamics of the system, including the synaptic weights in our model, an intuitive picture of RD and the effects of learning emerged. As illustrated in Fig.~\ref{fig:6}a, through exposure to a particular odor, synaptic plasticity mechanisms lead to a learned representation, $\mathcal{D}_0$, of a given odor in the PCx activity space. The representations of a particular odor are not unique -- they span a low-dimensional sub-manifold in the full representation space as illustrated in Fig.~\ref{fig:6}a. During the long time interval, $T$, between odor presentations, neural activity wanders off the sub-manifold due to stochastic synaptic dynamics to a point $\mathcal{D}_{T}$ away from it. However, the weight change induced by learning during  presentation of the odor at time $T\rightarrow T+\Delta t$  can drive the system back to another point, $\mathcal{D}_{T+\Delta t}$,  on (or near) the representation sub-manifold. Thus, the odor-dependent learning effectively reduces RD by suppressing fluctuations away from the low-dimensional odor-specific sub-manifold, i.e., $ \lVert \mathcal{D}_0 - \mathcal{D}_{T+\Delta t} \rVert < \lVert \mathcal{D}_0 - \mathcal{D}_T \rVert$ (see Fig. \ref{fig:6}a). 

To verify this intuitive picture, we measured the dynamics of relevant MTC-to-pyramidal weights in our model during an experiment where a particular familiar odor was presented every day after familiarization (Fig.~\ref{fig:5}a). As shown in Fig.~\ref{fig:6}b, odor presentation induces directed weight changes, i.e., changes with the same sign, while weights fluctuate randomly in-between test days. 

To characterize the collective weight changes in the whole system, we projected the dynamics of non-zero LOT connections onto the first three principal components that are obtained by applying PCA to the recorded weights during odor presentation on the first day (Fig.~\ref{fig:6}c). As we can see, while the random multiplicative fluctuations caused by the GMR process push the weights in all directions, STDP-mediated changes can compensate the random drift, consistently across one of the principal component directions (see arrows in Fig.~\ref{fig:6}c). 

Finally, to quantify the overall changes in weight space, we computed the normalized Euclidean distance with respect to the original set of weights using the first 10 principal components, which account for $\sim 99\%$ of the total variance:
\begin{equation}
\lVert \Delta J_{mtc}(t) \rVert = \dfrac{\lVert \mathbf{J}_{mtc}(t) - \mathbf{J}_{mtc}(0)\rVert}{\lVert\mathbf{J}_{mtc}(T) - \mathbf{J}_{mtc}(0)\rVert} \ ,
\end{equation}
where $\mathbf{J}_{mtc}(t)$ is a 10-dimensional projection of the LOT weights at time $t$, and $T$ represents the last time step after $9$ simulated days.  Fig.~\ref{fig:6}d shows that the Euclidean distance is reduced after each presentation of the stimulus (vertical dotted lines), which effectively decreases the rate at which drift takes place.

\begin{comment}
    \section*{A possible functional role of RD}\label{Section_5}

Notably, the idea of having ``learned'' representations that drift across time in the activity space, suggests a very interesting potential role of representational drift in odor discrimination. In particular, one can imagine that PCx responses to similar odors (i.e., converted at the OB level into strongly overlapping MTCs patterns of activity), will be initially mapped into nearby points on the representation manifold. If our picture is accurate, one would expect that those nearby representations will slowly drift in random directions due to the noisy weight dynamics, eventually increasing the separability of the responses to very similar odors. This is indeed what we observe when we presented a set of 12 unfamiliar but similar odors (meaning that MTCs responses to any pair of odors share 80\% of their characteristic latency times) to our original model (see Fig.~\ref{fig:6}). 
\end{comment}

\section*{Conclusions and Discussion}

Understanding how information from the external world is encoded in cortical neuronal activity requires careful analysis of experimental data in combination with computational modeling based on realistic neural networks. Here, we focus on the olfactory cortex and the experimentally observed  representational drift (RD) in the response of pyramidal cells, as reported by Schoonover {\emph et al.}~\cite{schoonover_representational_2021}. %aiming at understanding the origin, key characteristics, and possible functional significance of such a drift.

To understand the underlying mechanism for the observed RD phenomenon, we incorporate dynamics of synaptic plasticity in a realistic spiking neural network model of the mouse olfactory cortex. Our model reproduces all the main experimental findings reported in \cite{schoonover_representational_2021} quantitatively. More importantly, it elucidates a general mechanism underlying RD and the effects of learning, which may be applicable to other brain regions that exhibit RD.  
In what follows, we discuss our main findings and possible future directions. 

%\vspace{0.5cm}
\subsection*{Slow drift by stochastic weight variations} 

Synapses in the cortex are highly dynamic and can change in time regardless of the existence or absence of a stimulus, with synaptic efficiencies spontaneously fluctuating over a long timescale of days without altering their overall statistics.  As a result of this stochastic process, the response of cortical pyramidal cells  to a given odor signal—constituting the cortical representation of the odor—drifts slowly, while the statistical properties of the representation remain stable over time. Our study finds that the spontaneous weight fluctuations are best described by an Ornstein-Uhlenbeck process with multiplicative noise (Eq.~\ref{eq:GMR_process}), also known as the geometric mean reversion (GMR) process.
Our realistic network model, with weight dynamics governed by the GMR process, not only quantitatively reproduces all experimental results on RD in the piriform cortex \cite{schoonover_representational_2021} but also results in a log-normal distribution of the weights in steady state, consistent with empirical observations across brain regions ~\cite{loewenstein_predicting_2015, Buzsaki2014}. 

\subsection*{Learning suppresses representational drift} 
In the presence of odors that induce strong responses in the olfactory system, significant changes in synaptic weights can occur via local learning rules, such as spike-timing-dependent plasticity (STDP), during the relatively short time window of odor presentation. Rather than being random, these systematic weight changes drive the system towards a lower-dimensional sub-manifold in the representational space, as illustrated in Fig.~\ref{fig:6}a, effectively compensating for the deviations induced by the spontaneous noisy drift. Consequently, when the frequency of odor presentation is increased, drift rate is reduced, as illustrated in our model (Fig.~\ref{fig:5} and Fig.~\ref{fig:SI_FamVSUnf}f) and in agreement with experimental observations ~\cite{schoonover_representational_2021}. Notably, this mechanism of drift reduction requires the existence of a ``learned'' representation induced by previous exposure of the animal to a given odor (referred to as the ``familiarization" phase in the experiments~\cite{schoonover_representational_2021}). 

%Indeed, when we repeat the same numerical experiments with a high presentation frequency (daily) but without familiarization phase, we observe that, instead of being reduced, the amount of change with respect to the original representation is slightly higher than that with less frequent presentation (see Fig.S2xx in the SI). This apparent stimulus-induced ``drift" increase in the drift rate  is likely caused by the persistent changes of the neuron activities towards the representational sub-manifold driven by the frequent odor presentations since the system may start far away from the sub-manifold in the absence of pre-training.

%\vspace{0.35cm}
\subsection*{Absence of an invariant geometry} 
Consistent with the experiments by Schoonover et al~\cite{schoonover_representational_2021}, our model shows that the geometry of the drifting representation manifold is not invariant across time. Specifically, the relative angles between pairs of odor-specific responses in the representational manifold do not remain constant over time, both in the original experiments and our model.  
This observation contrasts with the findings in a recent study by Qin \emph{et al.} \cite{qin_coordinated_2023}, who developed a model to explain RD with Hebbian/anti-Hebbian networks. In their work, the learning dynamics aims to minimize the mismatch between the similarity of pairs of inputs and corresponding pairs of outputs, resulting in a coordinated drift that preserves the manifold geometry. 

On the functional side, we contend that geometric considerations are crucial when the inputs or stimuli to be encoded have a well-defined geometry. This is certainly the case for spatial locations or motion orientations encoded in the hippocampus, which is one of the focuses of study in \cite{qin_coordinated_2023}. In these scenarios, the representation manifold must reflect the actual spatial organization of stimuli, and this objective is best achieved if the representation manifold preserves geometric properties such as relative angles. Instead, in our modeling study of the piriform cortex presented here, the set of possible stimuli, i.e., odors, lacks any specific organization. As long as they are separable, the relative angles between representations of different odors do not seem to encode useful information; in particular, the similarity between two different odors may not be directly related to the scalar product (an Euclidean distance measure) in either the input space or the representational space.  

Indeed, studies by the Sharpee group \cite{Sharpee2018, Sharpee2019} have proposed that the olfactory space is hyperbolic rather than Euclidean, which could reflect a hierarchical organization in the odor space \cite{Krioukov}. Therefore, in future extensions of our work, we will seek to replace randomly organized odors with hierarchically structured ones to analyze whether the resulting representation manifold exhibits hyperbolic geometry (by keeping, e.g., hyperbolic inner products and angles fixed).

%\vspace{0.35cm}
\subsection*{Invariant properties of the representation manifold}
Even though the geometry of the representation manifold is not invariant, certain aspects of the representation remain preserved. In particular, the covariance matrix for pyramidal neurons presents an underlying block structure that is preserved over time. While the identity of the neurons participating in each block changes across days, the number of strongly correlated blocks remains constant and matches approximately the number of odors presented to the system (Fig.~\ref{fig:SI_DriftMeasures}k). Notably, a similar phenomenology was found by Kossio et al. in a work that studied RD using an associative memory computational model \cite{kossio_drifting_2021}. In particular, the authors proposed a mechanism by which STDP and homeostasis could maintain a constant representational structure despite the existence of drifting neural assemblies. However, unlike in our model, the emergence of this neuronal assemblies required a mechanism to generate symmetric connectivity matrices, which also encoded the block structure within their architecture \cite{kossio_drifting_2021}.

Besides the invariant representational structure in correlations, a principal component analysis of the pyramidal activity in response to the presented stimuli shows that the effective dimensionality of the representation manifold remains constant and is close to the number of encoded odors (Fig.\ref{fig:SI_DriftMeasures}l). We speculate that for hierarchically structured odor stimuli, the covariance matrix should exhibit a nested hierarchical block structure and, consequently, could be describable by an invariant hyperbolic geometry \cite{Krioukov,Sharpee2019}.

\subsection*{Beyond the olfactory cortex: shedding light on the slow vs fast drift conundrum}

Experimental studies have shown the emergence of drifting representations on very long timescales of days or weeks in diverse brain regions \cite{schoonover_representational_2021, Churchland2010, ziv_long-term_2013, driscoll_dynamic_2017}. However, %in manifest opposition to the generality of the above findings,
%other works found that experience, more than the passage of time, the driving force behind representational drift   \cite{khatib_active_2023,jacobson_experience-dependent_2018}. 
%In particular, 
in a recent work by Khatib \emph{et al.} \cite{khatib_active_2023}, where mice were trained to navigate a familiar maze, it was found that for a set amount of time, the more frequently the mice explored the environment, the greater the degree of drift observed in the neural representation of the spatial location in dorsal CA1 of the mouse hippocampus, a result that has been recently addressed computationally \cite{ratzon_representational_2024}.

We would like to point out that, contrary to what it might seem, this result is not at odds with the empirical observation made by Schoonover \emph{et al.} regarding drift slowing down with the frequency of stimulus presentation. Indeed, for the experiments in \cite{khatib_active_2023}, drift was measured between population neuron activities recorded with a time difference of $\sim3$ hours. Following our hypotheses, at these shorter timescales changes in representations are not really the result of a \emph{drift} as caused by the noisy weight dynamics, but actually originate from what we called a \emph{learning force} through STDP effects while mice were traversing the familiar environment. This conclusion is indeed supported by the findings of two other experimental papers in  the dorsal CA1 region of the mice hippocampus \cite{geva_time_2023}, and the telencephalic area Dp of the adult zebra fish \cite{jacobson_experience-dependent_2018}, which we discuss below.

In the Geva \emph{et al.} \cite{geva_time_2023} study, the authors carried out an experiment similar to the one in \cite{khatib_active_2023}, but sampled neural activity for a time-span of $3$ weeks. Within such much-longer recordings, they were able to confirm the existence of time-dependent drift that mostly affected changes in activity rates, and an experience-dependent drift that affected neural tuning curves. In another study, Jacobson \emph{et al.} \cite{jacobson_experience-dependent_2018} showed that the substantial variability observed in  zebra-fish Dp neural activity for a given odor across trials, which they called \emph{representation shift}, was severely reduced by an NMDA receptor antagonist, implying that these modifications were indeed experience-dependent.

%Notably, while overall population correlation decreased at a slower rate for the case of familiar representations in simulations of Cohort A mice (Fig. SX), we observed that tuning curve correlations (which capture the change in pyramidal units tuning to odors), decreased indeed faster for familiar odors presented every day, 

All these findings suggest that representational drift, as measured between population responses to the same input across days, is caused by two contributing mechanisms: (i) an actual random drift, which we hypothesize stems from noisy multiplicative weight dynamics; and (ii) a learning force, induced by STDP on a shorter timescale when the represented external input is experienced (e.g., positions in space \cite{ziv_long-term_2013, khatib_active_2023, geva_time_2023}) or presented (e.g., odors \cite{jacobson_experience-dependent_2018, schoonover_representational_2021} or visual stimuli \cite{marks_stimulus-dependent_2021, deitch_representational_2021}). While the first mechanism always leads to random drift, the effects of learning on RD depend on the timescale of measuring representational drift. On longer timescales (of days or weeks), learning leads to suppression of RD as it effectively reduces the effects of random drift. However, for shorter measurement timescales, learning can increase the measured drift as it drives the system deterministically towards its learned representation sub-manifold. The role of learning and its dependence on the measurement time scale are demonstrated analytically in a toy model in SI (Supplementary Note \ref{toymodel}).       

Although the main focus of this article was on reproducing and explaining through a biologically realistic computational model the experimental findings of Schoonover \emph{et al.} in \cite{schoonover_representational_2021}, preliminary results suggest that our proposed mechanisms for drift are also in very good agreement with experimental observations in other regions, such as a non-trivial alignment of drift changes with the directions of noise variance in the mouse posterior parietal cortex \cite{rule_stable_2020} (see SI, Supplementary Note \ref{NoiseAlignment}).% or the sparsification of neural coding in the dorsal CA1 region \cite{khatib_active_2023, ratzon_representational_2024}.

%In summary,} we hope this work will inspire new analyses to fully elucidate the causes and consequences of representational drift in specific brain regions.

\section*{Methods}

\subsection*{OB dynamics}\label{Methods:OB_Dynamics}

In our simulations, and following the model by Stern et al. \cite{stern_transformation_2018}, all MTCs belonging to glomerulus $i$ share a specific onset latency time, $\tilde{\tau}_{i}^{o}$, which is different for each odor, $o$. In particular, these latency times were randomly drawn from a uniform distribution such that only an average $10\%$ of all MTCs become responsive to a given odor within the inhalation period (i.e., $\tilde{\tau}_{i}^{o}<\tau_{inh}$).

Thus, while MTCs have a baseline firing rate at  $1.5\si{Hz}$, if glomerulus $i$ is activated during inhalation of a given odor, $o$, the instantaneous firing rate of all units belonging to this glomerulus jumps to $100\si{Hz}$ at time $\tilde{\tau}_{i}^{o}$, and then decays exponentially to the baseline rate with a characteristic time constant $\tau_{mtc}=50\si{ms}$.

For all simulations, one trial of odor presentation lasted $8$ respiration cycles, each consisting of a $200\si{ms}$ inhalation period, followed by a $300\si{ms}$ exhalation period. Figs.~\ref{fig:1}-\ref{fig:4}, Fig.~\ref{fig:SI_DriftMeasures} and Fig.~\ref{fig:SI_FamVSUnf}g: $N_{MTC}=2250$ and  $n_{glom}=90$, so that each glomerulus has $25$ associated MTCs. 

For simulations involving familiar and unfamiliar odors (Figs.~\ref{fig:5}-\ref{fig:6} and Fig.~\ref{fig:SI_FamVSUnf}a-f), we set $N_{MTC}=300$ and  $n_{glom}=60$, so that each glomerulus has $5$ associated MTCs. In this latter case and to avoid undesired cross-stimuli effects due to the network limited size (i.e., the familiarization process to one odor significantly affecting the representation of a second odor), we conducted simulations presenting one odor at a time, resetting to the exact same initial conditions for the network at the beginning of each experiment.

\subsection*{PCx dynamics}\label{Methods:PCx_Dynamics}

The below-threshold voltage dynamics of pyramidal neurons and  FBINs take the general form of a LIF equation:
\begin{equation}\label{eq:ch4_dvdt}
    \tau_m \dfrac{dV_i}{dt} = - \left(V_i(t) - V_{rest} \right) + I_{i}^{tot}(t)\ ,
\end{equation}
where $I_{i}^{tot}(t) = I_{i}^{exc}(t) + I_{i}^{inh}(t)$ is the sum of all incoming excitatory and inhibitory currents to neuron $i$, $\tau_m=15\si{ms}$ is the membrane characteristic time scale and $V_{rest}=-65\si{mV}$ is the resting potential. In the following equations, $J_{ij}^{XY}$ denote the synaptic efficiencies between presynaptic neuron $j$ belonging to population $X$ and postsynaptic neuron $i$  belonging to population $Y$ (m:MTC, p:pyramidal, f:FFIN, fb:FBIN). Defining $\Lambda_{j} = \sum_{f}\delta(t-t_j^f)$ as the train of spikes fired by neuron $j$, one can write for pyramidal units:
\begin{align}
    \tau_{exc}\dfrac{dI_{i}^{exc}}{dt} &= -I_{i}^{exc} + \sum_{j=1}^{N_{MTC}} J_{ij}^{mp}\Lambda_{j} + \sum_{j=1}^{N_{PYR}}J_{ij}^{pp}\Lambda_{j} \ ,\\
     \tau_{inh}\dfrac{dI_{i}^{inh}}{dt} &= -I_{i}^{inh} + \sum_{j=1}^{N_{FBIN}}J_{ij}^{fbp}\Lambda_{j} + \sum_{j=1}^{N_{FFIN}}J_{ij}^{fp}\Lambda_{j} \ .
\end{align}
Similarly for FBINs:
\begin{align}
    \tau_{exc}\dfrac{dI_{i}^{exc}}{dt} &= -I_{i}^{exc} + \sum_{j=1}^{N_{PYR}}J_{ij}^{pfb}\Lambda_{j} \ ,\\
    \tau_{inh}\dfrac{dI_{i}^{inh}}{dt} &= -I_{i}^{inh} + \sum_{j=1}^{N_{FBIN}}J_{ij}^{fbfb}\Lambda_{j} \ ,    
\end{align}
whereas for FFINs:
\begin{align}
    \tau_{exc}\dfrac{dI_{i}^{exc}}{dt} &= -I_{i}^{exc} + \sum_{j=1}^{N_{MTC}}J_{ij}^{mf}\Lambda_{j}\ ,\\
     \tau_{inh}\dfrac{dI_{i}^{inh}}{dt} &= -I_{i}^{inh} + \sum_{j=1}^{N_{FFIN}}J_{ij}^{ff}\Lambda_{j} \ .
\end{align}
In the above equations, $\tau_{exc}=20\si{ms}$ and $\tau_{inh}=20\si{ms}$ represent the characteristic decay time for the excitatory and inhibitory input currents. Notice that, for simplicity, we reabsorbed the membrane conductance $g_m$ into the definition of input current, so that all currents and synaptic efficiencies are expressed in units of voltage. Moreover, while in the original work by Stern et \emph{al.} all synaptic weights connecting the same type of neurons were set to a common value \citep{stern_transformation_2018}, in our model, weights were randomly drawn from a lognormal probability distribution with a given average $\langle J^* \rangle$ and standard deviation $\sigma= \langle J^* \rangle /2$.  

Once a neuron reaches its firing threshold, $V_{th}=-50\si{mV}$, its membrane potential is reset and clamped to a value $V_{reset}=-65\si{mV}$ for a refractory period,  $\tau_{ref}=1\si{ms}$, before it can evolve again according to \mbox{\eqref{eq:ch4_dvdt}}. We did not allow membrane potentials to decrease below a minimum value, $V_{min}=-75\si{mV}$. For all types of neurons, the dynamical equation was integrated using a 4th-order Runge-Kutta algorithm with a time step $\Delta t=0.0005s$.

For Figs.~\ref{fig:1}-\ref{fig:4}, Fig.~\ref{fig:SI_DriftMeasures} and Fig.~\ref{fig:SI_FamVSUnf}g, weight distribution averages and density of connections were chosen to enforce the same excitatory and inhibitory currents to each population as in \citep{stern_transformation_2018}: $\langle J^*_{mp} \rangle = \langle J^*_{mf} \rangle = 4\si{mV}$,  $\langle J^*_{fp} \rangle = \langle J^*_{ff} \rangle = 3\si{mV}, \langle J^*_{pp} \rangle = 1\si{mV}, \langle J^*_{pfb}  \rangle = 4\si{mV}, = \langle J^*_{fbp} \rangle = \langle J^*_{fbfb} \rangle = 3\si{mV}$. Density of connections: $p_{mp}=p_{mf}=0.022$, $p_{fp}=p_{ff}=0.4$, $p_{pp}=p_{pfb}=p_{fbp}=0.1$, $p_{fbfb}=0.065$.

For experiments involving familiar and unfamiliar odors (Figs.~\ref{fig:5}-\ref{fig:6} and Fig.~\ref{fig:SI_FamVSUnf}a-f), we reduced the number of neurons in the piriform cortex to $N_{PYR}=100$ and $N_{FBIN}=20$, limiting the inhibition of pyramidal cell activity to FBINs only, thus disregarding the effect of FFINs, which have been shown to simply modulate the amplitude but not the shape of pyramidal responses \cite{stern_transformation_2018}.  For the average weights:  $\langle J^*_{mp} \rangle = 3\si{mV}$,  $\langle J^*_{pp} \rangle = 5\si{mV}$, $\langle J^*_{pfb} \rangle = \langle J^*_{fbp} \rangle = \langle J^*_{fbfb} \rangle = 20\si{mV}$. Density of connections: $p_{mp}=0.025$,  $p_{pp}=0.1$, $p_{fbp}=p_{pfb}=p_{fbfb}=0.4$.

\subsection*{Weight dynamics} There are two different synaptic plasticity in our model, i.e., the slow stochastic weight changes described by the geometric mean-reversion process and the fast weight dynamics due to learning.

\subsubsection*{The geometric mean-reversion process}
The dynamics of synaptic weights in both, mitral-to-pyramidal and pyramidal-to-pyramidal connections, were modeled as a geometric mean-reversion process:
\begin{equation}
    \dot{J_k} = \omega (\mu - J_k(t))+ \sigma J_k(t) \xi_k(t),
\end{equation}
where $k$ indexes all non-zero synaptic efficiencies in the weight matrix, $\omega$ and $\sigma$ are constants for the drift and diffusion terms, $\mu=\langle J^* \rangle$ is the average non-zero weight in the initial Gaussian distribution (and will be, likewise, the average weight of the resulting stationary lognormal distribution) and $\xi(t)$ is a zero-mean Gaussian noise, such that   $\langle \xi_k(t) \rangle = 0$  and $\langle \xi_i(t) \xi_j(t') \rangle = \delta_{ij}\delta(t-t')$. In all simulations, we chose $\omega=\num{5e-7}$ and $\sigma=\num{4.5e-4}$. Before the beginning of each experiment, we also let the weights evolve for $32$ days under the above rule to ensure that a stationary lognormal probability distribution had been reached (see Fig.\ref{fig:1}e).

\subsubsection*{Spike-timing dependent plasticity} 
It has been experimentally observed that long-term potentiation (LTP) and long-term depression (LTD) of synaptic weights depend on the exact timing of the pre- and postsynaptic spikes \citep{markram_regulation_1997, bi_synaptic_2001, buchanan_activity_2010}. LTP is typically induced when the presynaptic spike precedes the postsynaptic one by an interval of $10$ to $20\si{ms}$, whereas LTD occurs if the order of spikes is reversed (see Fig.\ref{fig:1}a, bottom). This mechanism of STDP has been largely studied from a theoretical point of view \citep{rossum_stable_2000, gilson_stability_2011, gilson_stdp_2011, carlson_biologically_2013, effenberger_self-organization_2015} and many models have been proposed to investigate its functional implications (see \cite{caporale_spike_2008} and \cite{morrison_phenomenological_2008} for reviews on the topic). 

Mathematically speaking, the change in the synaptic weight induced by pre- and postsynaptic spikes at time $t_{pre}$ and $t_{post}$, respectively, can be written as:
\begin{equation} \label{eq:Ch4_GeneralSTDP}
    \Delta J = \Gamma (J; t_{post}-t_{pre})
\end{equation}
where $\Gamma(w; t_{post}-t_{pre})$ is the plasticity window, which can lead to potentiation (LTP) or depression (LTD) depending on the relative timing of the spikes, $\Delta t^*=t_{post}-t_{pre}$:
\begin{equation} \label{eq:ch4_wupdates}
\Gamma(J; \Delta t^*) =\begin{cases}
			f_{+}(J)\exp\left( -\dfrac{\lvert \Delta t^*\rvert}{\tau_{+}}\right), & \text{if $t_{pre}<t_{post}$}\\
            f_{-}(J)\exp\left( -\dfrac{\lvert \Delta t^*\rvert}{\tau_{-}}\right), & \text{if $t_{pre}>t_{post}$} \ . 
		 \end{cases}
\end{equation}
Within the above expression different choices of the scaling functions for potentiation,  $f_{+}(w)$, and depression, $f_{-}(w)$, can give rise to different models of STDP \citep{morrison_phenomenological_2008}. 
Here we use a \emph{multiplicative} STDP, as originally proposed in \cite{rossum_stable_2000} ---on the basis of experimental observations in \citep{bi_synaptic_1998}---, for which the LTP and LTD scaling functions read:
\begin{align}
      f_{+}(J) & = a_{+} \ , \\
      f_{-}(J) & = -a_{-}J \ ,
\end{align}
for some constant values $a_{+}$ and $a_{-}$. All simulations were ran using $\tau_+=17\si{ms}$ and $\tau_-=34\si{ms}$ for the LTP and LTD windows, respectively, and $a_+=0.0005\langle J^{*} \rangle$. The gain factor for LTD, $a_{-}$, was chosen in each case so that the average weight of the stationary distribution under the STDP rule ($\langle J^{st} \rangle = (a_+ \tau_+) / (a_- \tau_-$), see  \citep{gilson_stability_2011}), matches the average weight,  $\langle J^{*} \rangle$, of the expected lognormal distribution under the GMR process. 

\subsection*{Measures of drifting representations}

For each day $d$, odor $o$ and trial $m$, the representation of an odor, $\mathbf{x}_{d,o,m}$, was computed by averaging pyramidal responses in time across four $2\si{s}$-windows after odor onset, then concatenating the corresponding vectors so that $\mathbf{x}_{d,o,m}\in\mathbb{R}^{4N_{PYR}}$. For each neuron (i.e., element in $\mathbf{x}_{d,o,m}$) its average baseline spontaneous rate, computed as the average rate across all days during the transient periods, was subtracted in all cases before computing any drift-related quantity. 

Correlation between same-odor representations at days $p$ and $q$, with $p \neq q$, was defined as the average across odors of the Pearson's correlation coefficient between trial-averaged population vectors at the corresponding days:
\begin{equation} \label{eq:pearson_corr}
    c_{p,q} = \dfrac{1}{n_{odors}} \sum_{o=1}^{n_{odors}} \dfrac{\langle(\mathbf{x}_{p,o}-\overline{x}_{p,o})(\mathbf{x}_{q,o} -\overline{x}_{q,o})\rangle}{\sigma_{\mathbf{x}_{p,o}}\sigma_{\mathbf{x}_{q,o}}}
\end{equation}
where $\mathbf{x}_{p,o}=M^{-1}\sum_{m=1}^{M} \mathbf{x}_{p,o,m}$ is the trial-averaged population response to odor $o$  on day $p$, and $\overline{x}_{p,o}$ and $\sigma_{\mathbf{x}_{p,o}}$ define its mean and standard deviation, respectively. 

Trivially, one can then define within-day correlations between odor responses by averaging population vectors across even and odd trials separately: 
\begin{equation} \label{eq:SI_Ch4_WithinDayCorr}
    c_{p} = \dfrac{1}{n_{odors}} \sum_{o=1}^{n_{odors}} \dfrac{\langle(\mathbf{x}_{p,o}^{even}-\overline{x}_{p,o}^{even})(\mathbf{x}_{p,o}^{odd} -\overline{x}_{p,o}^{odd})\rangle}{\sigma_{\mathbf{x}_{p,o}^{even}}\sigma_{\mathbf{x}_{p,o}^{odd}}} \ ,
\end{equation}
where $\mathbf{x}_{p,o}^{even}$ ( $\mathbf{x}_{p,o}^{even}$) is the population response to odor $o$  on day $p$ averaged over all even (odd) trials. 

On the other hand, the average correlation between same-odor responses separated by a time interval of $\Delta$-days, is:
\begin{equation}
    \overline{c}_\Delta = \frac{1}{n_{\Delta}}\sum_{p,q \ : \ \lvert p-q\rvert=\Delta} c_{p,q} \ , 
\end{equation}
where $n_{\Delta}$ is the number of pairs of test days separated by a time interval $\Delta$. 

%Pero esto tiene sentido si hemos llegado a un regime estacionario donde no depende de p y q sepradamente sino de la diferencia, right? 
%---> Sí, tienes razón, supongo que es una asumption del experimento!

Similarly, the average angle between a pair of population vectors representing the same odor at days $p$ and $q$ can be written as:
\begin{equation}
    \theta_{p,q} =\frac{1}{n_{\Delta}} \sum_{o=1}^{n_{odors}}\theta_{p,q}^{o} =  \dfrac{1}{n_{odors}} \sum_{o=1}^{n_{odors}}\cos^{-1} \left( \dfrac{\mathbf{x}_{p,o} \cdot \mathbf{x}_{q,o}}{\|\mathbf{x}_{p,o}\|\|\mathbf{x}_{q,o}\|} \right) \ .
\end{equation}
and the average within-day angle at each day $p$ is defined as:
\begin{equation} \label{eq:SI_Ch4_WithinDayAngle}
    \theta_p = \dfrac{1}{n_{odors}}\sum_{o=1}^{n_{odors}} \cos^{-1} \dfrac{\mathbf{x}_{p,o}^{even} \cdot \mathbf{x}_{p,o}^{odd}}{\|\mathbf{x}_{p,o}^{even}\|\|\mathbf{x}_{p,o}^{odd}\|} \ . 
\end{equation}
To correct for within-day variability in the angle between representations of the same odor, we followed \cite{schoonover_representational_2021} and computed the average within-day angle, $\overline{\theta}=\frac{1}{n_{days}}\sum_{p}\theta_{p}$ , between same-odor, same-day population responses. Thus, the average corrected angle between any two representations measured on tests separated by $\Delta$-days was finally computed as: 
\begin{equation}\label{eq:ch4_AvCorrectedAngle}
    \overline{\theta}_\Delta = n_{\Delta}^{-1}\sum_{p,q \ : \ \lvert p-q\rvert=\Delta} (\theta_{p,q} - \overline{\theta}) \ .
\end{equation}
Using the above quantity, it is possible to measure the rate of drift (in angles per day and corrected for within-day fluctuations) as:
\begin{equation} \label{eq:Ch4_DriftRate}
    \overline{r}= \Biggl\langle \dfrac{ \overline{\theta}_\Delta}{\Delta}\Biggr\rangle_\Delta \ ,
\end{equation}
where the average is taken across  all possible time intervals, $\Delta$, between any pair of test days.

\begin{comment}
   \subsection*{Classifying odors from PCx responses} \label{Methods:Classification}

Following the methods in \cite{schoonover_representational_2021}, we trained a Support Vector Machine (SVM) with linear kernel and L2-regularization to classify the odors from the population responses of pyramidal neurons in the PCx. For within-day classification, we used leave-one-out cross-validation, training on all but one of the 56 trials on a given test day (8 odors, 7 trials) and then testing on the trial that was left out. This procedure was repeated until all trials on a given day had been tested in this way. For across-day classification, the model was trained using the 56 trials on one day, and then tested on all 56 trials on another day. As control case, we measured performance on shuffled data by randomly permuting the odor stimulus labels on the test set. Notably, in the experimental results of Schoonover \emph{et al}, training and testing of SVM was limited to the lowest number of stable single units for any across-day comparison and mouse (41 units). To produce a sensible comparison, we randomly selected the same number of units in our simulations to train and test the SVM, mimicking the heavily subsampled regime in which the experiments take place.  
\end{comment}

\subsection*{Measures of drifting geometry} \label{Methods:DriftingGeometry}

Matrix dissimilarity between days $p$ and $q$  was taken as:
\begin{equation}
   \lVert A^{p,q}  \rVert_{F} \coloneqq \lVert A^p - A^q  \rVert_{F} = \sqrt{\sum_{k=1}^{M}\sum_{i=1}^{n_{odors}} \lvert a_{k,i}^p - a_{k,i}^q\rvert^2} \ , 
\end{equation}
where $\lVert A^{p,q}  \rVert_{F} = 0$ for identical matrices. Given the above measure, a corrected matrix dissimilarity was computed as:
\begin{equation} \label{eq:SI_Ch4_CoorrectedDissimilarity}
   \lVert \hat{A}^{p,q}  \rVert_{F} \coloneqq \dfrac{\lVert A^{p,q}  \rVert_{F} - \overline{\lVert A^{w} \rVert}_{F}}{\lVert A^{s} \rVert_{F} - \overline{\lVert A^{w} \rVert}_{F}} \ ,
\end{equation}
where $\overline{\lVert A^{w} \rVert}_{F}\coloneqq (1/n_{days})\sum_{p=1}^{n_{days}} \lVert A^{p}_{odd} - A^{p}_{even} \rVert$ is the mean across all days of the within-day Frobenius norm between similarity matrices computed in odd and even trials, and $\lVert A^{s} \rVert_{F}$ is the Frobenius norm between similarity matrices measured on the first and last day of the experiment after odors shuffling. In this way, $ \lVert \hat{A}^{p,q}  \rVert_{F}$ is (on average) bounded between zero, for angle drifts on the order of intra-day fluctuations, and one, for the shuffled case. 

\subsection*{Population statistics}\label{Methods:PopStatistics} 

To identify responsive units, on each day and for each odor a Wilcoxon rank-sum test \citep{haynes_wilcoxon_2013} was performed between the spike count during the 4 seconds before stimulus onset on all trials and the spike count on all trials during the odor presentation, using a significance level of $\alpha=0.005$. On each day, the number of responsive neurons was averaged over all presented stimuli and normalized by population size to compute the average fraction of responsive neurons (see Fig. \ref{fig:2}e (left panel)).

Given spontaneous baseline-subtracted responses, $r_{j,o}$, for each unit, $j$,  to a given odor, $o$, average population sparseness was defined as: 
\begin{equation} \label{eq:SI_Ch4_PopSparseness}
    S_p = \dfrac{N - 1}{n_{odors}N} \sum_{o=1}^{n_{odors}} \left(1 - \dfrac{\left(N^{-1}\sum_{j=1}^N r_{j,o} \right)^2}{N^{-1}\sum_{j=1}^N r_{j,o}^2}\right) \ ,
\end{equation}
whereas the average lifetime sparseness across all units was given by:
\begin{equation}\label{eq:SI_Ch4_LTSparseness}
    S_{lt} = \dfrac{n_{odors} - 1}{n_{odors}N} \sum_{j=1}^{N} \left(1 - \dfrac{\left(n_{odors}^{-1}\sum_{o=1}^{n_{odors}} r_{j,o} \right)^2}{n_{odors}^{-1}\sum_{o=1}^{n_{odors}} r_{j,o}^2}\right) \ .
\end{equation}

\begin{acknowledgements}

This work has been supported by Grant No.
PID2023-149174NB-I00 financed by MICIU/AEI/10.13039/501100011033 and EDRF/EU funds, as well as by
PID2020-113681GB-I00, MICIU/AEI/10.13039/501100011033. 
\end{acknowledgements}

\section*{Bibliography}
\bibliography{RepDrift_BIB}

\begin{thebibliography}{56}
\providecommand{\natexlab}[1]{#1}
\providecommand{\url}[1]{\texttt{#1}}
\expandafter\ifx\csname urlstyle\endcsname\relax
  \providecommand{\doi}[1]{doi: #1}\else
  \providecommand{\doi}{doi: \begingroup \urlstyle{rm}\Url}\fi

\bibitem[Chung and Abbott(2021)]{chung_neural_2021}
SueYeon Chung and L.~F. Abbott.
\newblock Neural population geometry: {An} approach for understanding biological and artificial neural networks.
\newblock \emph{Current Opinion in Neurobiology}, 70:\penalty0 137--144, October 2021.
\newblock ISSN 0959-4388.
\newblock \doi{10.1016/j.conb.2021.10.010}.

\bibitem[Stringer et~al.(2019)Stringer, Pachitariu, Steinmetz, Carandini, and Harris]{Stringer2019}
Carsen Stringer, Marius Pachitariu, Nicholas Steinmetz, Matteo Carandini, and Kenneth~D Harris.
\newblock High-dimensional geometry of population responses in visual cortex.
\newblock \emph{Nature}, 571\penalty0 (7765):\penalty0 361--365, 2019.

\bibitem[Manley et~al.(2024)Manley, Lu, Barber, Demas, Kim, Meyer, Traub, and Vaziri]{Manley-1million}
Jason Manley, Sihao Lu, Kevin Barber, Jeffrey Demas, Hyewon Kim, David Meyer, Francisca~Mart{\'\i}nez Traub, and Alipasha Vaziri.
\newblock Simultaneous, cortex-wide dynamics of up to 1 million neurons reveal unbounded scaling of dimensionality with neuron number.
\newblock \emph{Neuron}, 112\penalty0 (10):\penalty0 1694--1709, 2024.

\bibitem[Abbott and Sejnowski(1999)]{Abbott1999}
Laurence~F Abbott and Terrence~Joseph Sejnowski.
\newblock \emph{Neural codes and distributed representations: foundations of neural computation}.
\newblock Mit Press, 1999.

\bibitem[Yuste(2015)]{Yuste2015}
Rafael Yuste.
\newblock From the neuron doctrine to neural networks.
\newblock \emph{Nature reviews neuroscience}, 16\penalty0 (8):\penalty0 487--497, 2015.

\bibitem[Schoonover et~al.(2021)Schoonover, Ohashi, Axel, and Fink]{schoonover_representational_2021}
Carl~E. Schoonover, Sarah~N. Ohashi, Richard Axel, and Andrew J.~P. Fink.
\newblock Representational drift in primary olfactory cortex.
\newblock \emph{Nature}, 594\penalty0 (7864):\penalty0 541--546, June 2021.
\newblock ISSN 1476-4687.
\newblock \doi{10.1038/s41586-021-03628-7}.
\newblock Bandiera\_abtest: a Cg\_type: Nature Research Journals Number: 7864 Primary\_atype: Research Publisher: Nature Publishing Group Subject\_term: Cortex;Olfactory cortex;Sensory processing Subject\_term\_id: cortex;olfactory-cortex;sensory-processing.

\bibitem[Driscoll et~al.(2017)Driscoll, Pettit, Minderer, Chettih, and Harvey]{driscoll_dynamic_2017}
Laura~N. Driscoll, Noah~L. Pettit, Matthias Minderer, Selmaan~N. Chettih, and Christopher~D. Harvey.
\newblock Dynamic {Reorganization} of {Neuronal} {Activity} {Patterns} in {Parietal} {Cortex}.
\newblock \emph{Cell}, 170\penalty0 (5):\penalty0 986--999.e16, August 2017.
\newblock ISSN 0092-8674, 1097-4172.
\newblock \doi{10.1016/j.cell.2017.07.021}.
\newblock Publisher: Elsevier.

\bibitem[Khatib et~al.(2022)Khatib, Ratzon, Sellevoll, Barak, Morris, and Derdikman]{khatib_experience_2022}
Dorgham Khatib, Aviv Ratzon, Mariell Sellevoll, Omri Barak, Genela Morris, and Dori Derdikman.
\newblock Experience, not time, determines representational drift in the hippocampus, August 2022.
\newblock Pages: 2022.08.31.506041 Section: New Results.

\bibitem[Ratzon et~al.(2024)Ratzon, Derdikman, and Barak]{ratzon_representational_2024}
Aviv Ratzon, Dori Derdikman, and Omri Barak.
\newblock Representational drift as a result of implicit regularization.
\newblock \emph{eLife}, 12, April 2024.
\newblock \doi{10.7554/eLife.90069.2}.
\newblock Publisher: eLife Sciences Publications Limited.

\bibitem[Marks and Goard(2021)]{marks_stimulus-dependent_2021}
Tyler~D. Marks and Michael~J. Goard.
\newblock Stimulus-dependent representational drift in primary visual cortex.
\newblock \emph{Nature Communications}, 12\penalty0 (1):\penalty0 5169, August 2021.
\newblock ISSN 2041-1723.
\newblock \doi{10.1038/s41467-021-25436-3}.
\newblock Bandiera\_abtest: a Cc\_license\_type: cc\_by Cg\_type: Nature Research Journals Number: 1 Primary\_atype: Research Publisher: Nature Publishing Group Subject\_term: Neural circuits;Sensory processing Subject\_term\_id: neural-circuit;sensory-processing.

\bibitem[Kossio et~al.(2021)Kossio, Goedeke, Klos, and Memmesheimer]{kossio_drifting_2021}
Yaroslav Felipe~Kalle Kossio, Sven Goedeke, Christian Klos, and Raoul-Martin Memmesheimer.
\newblock Drifting assemblies for persistent memory: {Neuron} transitions and unsupervised compensation.
\newblock \emph{Proceedings of the National Academy of Sciences}, 118\penalty0 (46), November 2021.
\newblock ISSN 0027-8424, 1091-6490.
\newblock \doi{10.1073/pnas.2023832118}.
\newblock Publisher: National Academy of Sciences Section: Biological Sciences.

\bibitem[Qin et~al.(2023)Qin, Farashahi, Lipshutz, Sengupta, Chklovskii, and Pehlevan]{qin_coordinated_2023}
Shanshan Qin, Shiva Farashahi, David Lipshutz, Anirvan~M. Sengupta, Dmitri~B. Chklovskii, and Cengiz Pehlevan.
\newblock Coordinated drift of receptive fields in {Hebbian}/anti-{Hebbian} network models during noisy representation learning.
\newblock \emph{Nature Neuroscience}, 26\penalty0 (2):\penalty0 339--349, February 2023.
\newblock ISSN 1546-1726.
\newblock \doi{10.1038/s41593-022-01225-z}.
\newblock Number: 2 Publisher: Nature Publishing Group.

\bibitem[Rule and O’Leary(2022)]{rule_self-healing_2022}
Michael~E. Rule and Timothy O’Leary.
\newblock Self-healing codes: {How} stable neural populations can track continually reconfiguring neural representations.
\newblock \emph{Proceedings of the National Academy of Sciences}, 119\penalty0 (7):\penalty0 e2106692119, February 2022.
\newblock \doi{10.1073/pnas.2106692119}.
\newblock Publisher: Proceedings of the National Academy of Sciences.

\bibitem[Aitken et~al.(2022)Aitken, Garrett, Olsen, and Mihalas]{aitken_geometry_2022}
Kyle Aitken, Marina Garrett, Shawn Olsen, and Stefan Mihalas.
\newblock The geometry of representational drift in natural and artificial neural networks.
\newblock \emph{PLOS Computational Biology}, 18\penalty0 (11):\penalty0 e1010716, November 2022.
\newblock ISSN 1553-7358.
\newblock \doi{10.1371/journal.pcbi.1010716}.
\newblock Publisher: Public Library of Science.

\bibitem[Delamare et~al.(2023)Delamare, Zaki, Cai, and Clopath]{delamare_drift_2023}
Geoffroy Delamare, Yosif Zaki, Denise~J. Cai, and Claudia Clopath.
\newblock Drift of neural ensembles driven by slow fluctuations of intrinsic excitability.
\newblock \emph{eLife}, 12, December 2023.
\newblock \doi{10.7554/eLife.88053.2}.
\newblock Publisher: eLife Sciences Publications Limited.

\bibitem[Blazing and Franks(2020)]{blazing_odor_2020}
Robin~M Blazing and Kevin~M Franks.
\newblock Odor coding in piriform cortex: mechanistic insights into distributed coding.
\newblock \emph{Current Opinion in Neurobiology}, 64:\penalty0 96--102, October 2020.
\newblock ISSN 0959-4388.
\newblock \doi{10.1016/j.conb.2020.03.001}.

\bibitem[Buck and Axel(1991)]{buck_novel_1991}
L.~Buck and R.~Axel.
\newblock A novel multigene family may encode odorant receptors: a molecular basis for odor recognition.
\newblock \emph{Cell}, 65\penalty0 (1):\penalty0 175--187, April 1991.
\newblock ISSN 0092-8674.
\newblock \doi{10.1016/0092-8674(91)90418-x}.

\bibitem[Malnic et~al.(1999)Malnic, Hirono, Sato, and Buck]{malnic_combinatorial_1999}
Bettina Malnic, Junzo Hirono, Takaaki Sato, and Linda~B. Buck.
\newblock Combinatorial {Receptor} {Codes} for {Odors}.
\newblock \emph{Cell}, 96\penalty0 (5):\penalty0 713--723, March 1999.
\newblock ISSN 0092-8674, 1097-4172.
\newblock \doi{10.1016/S0092-8674(00)80581-4}.
\newblock Publisher: Elsevier.

\bibitem[Jiang et~al.(2015)Jiang, Gong, Hu, Ni, Pasi, and Matsunami]{jiang_molecular_2015}
Yue Jiang, Naihua~Natalie Gong, Xiaoyang~Serene Hu, Mengjue~Jessica Ni, Radhika Pasi, and Hiroaki Matsunami.
\newblock Molecular profiling of activated olfactory neurons identifies odorant receptors for odors in vivo.
\newblock \emph{Nature Neuroscience}, 18\penalty0 (10):\penalty0 1446--1454, October 2015.
\newblock ISSN 1546-1726.
\newblock \doi{10.1038/nn.4104}.
\newblock Number: 10 Publisher: Nature Publishing Group.

\bibitem[Mombaerts et~al.(1996)Mombaerts, Wang, Dulac, Chao, Nemes, Mendelsohn, Edmondson, and Axel]{mombaerts_visualizing_1996}
Peter Mombaerts, Fan Wang, Catherine Dulac, Steve~K. Chao, Adriana Nemes, Monica Mendelsohn, James Edmondson, and Richard Axel.
\newblock Visualizing an {Olfactory} {Sensory} {Map}.
\newblock \emph{Cell}, 87\penalty0 (4):\penalty0 675--686, November 1996.
\newblock ISSN 0092-8674, 1097-4172.
\newblock \doi{10.1016/S0092-8674(00)81387-2}.
\newblock Publisher: Elsevier.

\bibitem[Hálasz and Greer(1993)]{halasz_terminal_1993}
Norbert Hálasz and Charles~A. Greer.
\newblock Terminal arborizations of olfactory nerve fibers in the glomeruli of the olfactory bulb.
\newblock \emph{Journal of Comparative Neurology}, 337\penalty0 (2):\penalty0 307--316, 1993.
\newblock ISSN 1096-9861.
\newblock \doi{10.1002/cne.903370211}.
\newblock \_eprint: https://onlinelibrary.wiley.com/doi/pdf/10.1002/cne.903370211.

\bibitem[Stern et~al.(2018)Stern, Bolding, Abbott, and Franks]{stern_transformation_2018}
Merav Stern, Kevin~A Bolding, LF~Abbott, and Kevin~M Franks.
\newblock A transformation from temporal to ensemble coding in a model of piriform cortex.
\newblock \emph{eLife}, 7:\penalty0 e34831, March 2018.
\newblock ISSN 2050-084X.
\newblock \doi{10.7554/eLife.34831}.
\newblock Publisher: eLife Sciences Publications, Ltd.

\bibitem[Bolding and Franks(2018)]{bolding_recurrent_2018}
Kevin~A. Bolding and Kevin~M. Franks.
\newblock Recurrent cortical circuits implement concentration-invariant odor coding.
\newblock \emph{Science}, 361\penalty0 (6407):\penalty0 eaat6904, September 2018.
\newblock \doi{10.1126/science.aat6904}.
\newblock Publisher: American Association for the Advancement of Science.

\bibitem[Bolding et~al.(2020)Bolding, Nagappan, Han, Wang, and Franks]{bolding_recurrent_2020}
Kevin~A Bolding, Shivathmihai Nagappan, Bao-Xia Han, Fan Wang, and Kevin~M Franks.
\newblock Recurrent circuitry is required to stabilize piriform cortex odor representations across brain states.
\newblock \emph{eLife}, 9:\penalty0 e53125, July 2020.
\newblock ISSN 2050-084X.
\newblock \doi{10.7554/eLife.53125}.
\newblock Publisher: eLife Sciences Publications, Ltd.

\bibitem[Meissner-Bernard et~al.(2023)Meissner-Bernard, Zenke, and Friedrich]{Meissner2023}
Claire Meissner-Bernard, Friedemann Zenke, and Rainer~W Friedrich.
\newblock Geometry and dynamics of representations in a precisely balanced memory network related to olfactory cortex.
\newblock \emph{bioRxiv}, pages 2023--12, 2023.

\bibitem[Wilson et~al.(2004)Wilson, Best, and Sullivan]{wilson_plasticity_2004}
D.~A. Wilson, A.~R. Best, and R.~M. Sullivan.
\newblock Plasticity in the {Olfactory} {System}: {Lessons} for the {Neurobiology} of {Memory}.
\newblock \emph{The Neuroscientist}, 10\penalty0 (6):\penalty0 513--524, December 2004.
\newblock ISSN 1073-8584.
\newblock \doi{10.1177/1073858404267048}.
\newblock Publisher: SAGE Publications Inc STM.

\bibitem[Ito et~al.(2008)Ito, Ong, Raman, and Stopfer]{ito_olfactory_2008}
Iori Ito, Rose Chik-ying Ong, Baranidharan Raman, and Mark Stopfer.
\newblock Olfactory learning and spike timing dependent plasticity.
\newblock \emph{Communicative \& Integrative Biology}, 1\penalty0 (2):\penalty0 170--171, 2008.
\newblock ISSN 1942-0889.

\bibitem[Ma et~al.(2012)Ma, Zhao, Cai, Zhang, Ren, Ji, Tian, and Lu]{ma_regulation_2012}
Teng-Fei Ma, Xiao-Lei Zhao, Lei Cai, Nan Zhang, Si-Qiang Ren, Fang Ji, Tian Tian, and Wei Lu.
\newblock Regulation of {Spike} {Timing}-{Dependent} {Plasticity} of {Olfactory} {Inputs} in {Mitral} {Cells} in the {Rat} {Olfactory} {Bulb}.
\newblock \emph{PLOS ONE}, 7\penalty0 (4):\penalty0 e35001, April 2012.
\newblock ISSN 1932-6203.
\newblock \doi{10.1371/journal.pone.0035001}.
\newblock Publisher: Public Library of Science.

\bibitem[Cohen et~al.(2015)Cohen, Wilson, and Barkai]{cohen_differential_2015}
Yaniv Cohen, Donald~A. Wilson, and Edi Barkai.
\newblock Differential {Modifications} of {Synaptic} {Weights} {During} {Odor} {Rule} {Learning}: {Dynamics} of {Interaction} {Between} the {Piriform} {Cortex} with {Lower} and {Higher} {Brain} {Areas}.
\newblock \emph{Cerebral Cortex}, 25\penalty0 (1):\penalty0 180--191, January 2015.
\newblock ISSN 1047-3211.
\newblock \doi{10.1093/cercor/bht215}.

\bibitem[Jacobson et~al.(2018)Jacobson, Rupprecht, and Friedrich]{jacobson_experience-dependent_2018}
Gilad~A. Jacobson, Peter Rupprecht, and Rainer~W. Friedrich.
\newblock Experience-{Dependent} {Plasticity} of {Odor} {Representations} in the {Telencephalon} of {Zebrafish}.
\newblock \emph{Current Biology}, 28\penalty0 (1):\penalty0 1--14.e3, January 2018.
\newblock ISSN 0960-9822.
\newblock \doi{10.1016/j.cub.2017.11.007}.
\newblock Publisher: Elsevier.

\bibitem[Kumar et~al.(2021)Kumar, Barkai, and Schiller]{kumar_plasticity_2021}
Amit Kumar, Edi Barkai, and Jackie Schiller.
\newblock Plasticity of olfactory bulb inputs mediated by dendritic {NMDA}-spikes in rodent piriform cortex.
\newblock \emph{eLife}, 10:\penalty0 e70383, October 2021.
\newblock ISSN 2050-084X.
\newblock \doi{10.7554/eLife.70383}.
\newblock Publisher: eLife Sciences Publications, Ltd.

\bibitem[Loewenstein et~al.(2011)Loewenstein, Kuras, and Rumpel]{loewenstein_multiplicative_2011}
Yonatan Loewenstein, Annerose Kuras, and Simon Rumpel.
\newblock Multiplicative {Dynamics} {Underlie} the {Emergence} of the {Log}-{Normal} {Distribution} of {Spine} {Sizes} in the {Neocortex} {In} {Vivo}.
\newblock \emph{Journal of Neuroscience}, 31\penalty0 (26):\penalty0 9481--9488, June 2011.
\newblock ISSN 0270-6474, 1529-2401.
\newblock \doi{10.1523/JNEUROSCI.6130-10.2011}.
\newblock Publisher: Society for Neuroscience Section: Articles.

\bibitem[Buzs{\'a}ki and Mizuseki(2014)]{Buzsaki2014}
Gy{\"o}rgy Buzs{\'a}ki and Kenji Mizuseki.
\newblock The log-dynamic brain: how skewed distributions affect network operations.
\newblock \emph{Nature Reviews Neuroscience}, 15\penalty0 (4):\penalty0 264--278, 2014.

\bibitem[Rossum et~al.(2000)Rossum, Bi, and Turrigiano]{rossum_stable_2000}
M.~C. W.~van Rossum, G.~Q. Bi, and G.~G. Turrigiano.
\newblock Stable {Hebbian} {Learning} from {Spike} {Timing}-{Dependent} {Plasticity}.
\newblock \emph{Journal of Neuroscience}, 20\penalty0 (23):\penalty0 8812--8821, December 2000.
\newblock ISSN 0270-6474, 1529-2401.
\newblock \doi{10.1523/JNEUROSCI.20-23-08812.2000}.
\newblock Publisher: Society for Neuroscience Section: ARTICLE.

\bibitem[Loewenstein et~al.(2015)Loewenstein, Yanover, and Rumpel]{loewenstein_predicting_2015}
Yonatan Loewenstein, Uri Yanover, and Simon Rumpel.
\newblock Predicting the {Dynamics} of {Network} {Connectivity} in the {Neocortex}.
\newblock \emph{Journal of Neuroscience}, 35\penalty0 (36):\penalty0 12535--12544, September 2015.
\newblock ISSN 0270-6474, 1529-2401.
\newblock \doi{10.1523/JNEUROSCI.2917-14.2015}.
\newblock Publisher: Society for Neuroscience Section: Articles.

\bibitem[Zhou et~al.(2018)Zhou, Smith, and Sharpee]{Sharpee2018}
Yuansheng Zhou, Brian~H Smith, and Tatyana~O Sharpee.
\newblock Hyperbolic geometry of the olfactory space.
\newblock \emph{Science advances}, 4\penalty0 (8):\penalty0 eaaq1458, 2018.

\bibitem[Sharpee(2019)]{Sharpee2019}
Tatyana~O Sharpee.
\newblock An argument for hyperbolic geometry in neural circuits.
\newblock \emph{Current opinion in neurobiology}, 58:\penalty0 101--104, 2019.

\bibitem[Krioukov et~al.(2010)Krioukov, Papadopoulos, Kitsak, Vahdat, and Bogun{\'a}]{Krioukov}
Dmitri Krioukov, Fragkiskos Papadopoulos, Maksim Kitsak, Amin Vahdat, and Mari{\'a}n Bogun{\'a}.
\newblock Hyperbolic geometry of complex networks.
\newblock \emph{Physical Review E—Statistical, Nonlinear, and Soft Matter Physics}, 82\penalty0 (3):\penalty0 036106, 2010.

\bibitem[Churchland et~al.(2010)Churchland, Yu, Cunningham, Sugrue, Cohen, Corrado, Newsome, Clark, Hosseini, Scott, et~al.]{Churchland2010}
Mark~M Churchland, Byron~M Yu, John~P Cunningham, Leo~P Sugrue, Marlene~R Cohen, Greg~S Corrado, William~T Newsome, Andrew~M Clark, Paymon Hosseini, Benjamin~B Scott, et~al.
\newblock Stimulus onset quenches neural variability: a widespread cortical phenomenon.
\newblock \emph{Nature neuroscience}, 13\penalty0 (3):\penalty0 369--378, 2010.

\bibitem[Ziv et~al.(2013)Ziv, Burns, Cocker, Hamel, Ghosh, Kitch, Gamal, and Schnitzer]{ziv_long-term_2013}
Yaniv Ziv, Laurie~D. Burns, Eric~D. Cocker, Elizabeth~O. Hamel, Kunal~K. Ghosh, Lacey~J. Kitch, Abbas~El Gamal, and Mark~J. Schnitzer.
\newblock Long-term dynamics of {CA1} hippocampal place codes.
\newblock \emph{Nature Neuroscience}, 16\penalty0 (3):\penalty0 264--266, March 2013.
\newblock ISSN 1546-1726.
\newblock \doi{10.1038/nn.3329}.
\newblock Number: 3 Publisher: Nature Publishing Group.

\bibitem[Khatib et~al.(2023)Khatib, Ratzon, Sellevoll, Barak, Morris, and Derdikman]{khatib_active_2023}
Dorgham Khatib, Aviv Ratzon, Mariell Sellevoll, Omri Barak, Genela Morris, and Dori Derdikman.
\newblock Active experience, not time, determines within-day representational drift in dorsal {CA1}.
\newblock \emph{Neuron}, 111\penalty0 (15):\penalty0 2348--2356.e4, August 2023.
\newblock ISSN 0896-6273.
\newblock \doi{10.1016/j.neuron.2023.05.014}.
\newblock Publisher: Elsevier.

\bibitem[Geva et~al.(2023)Geva, Deitch, Rubin, and Ziv]{geva_time_2023}
Nitzan Geva, Daniel Deitch, Alon Rubin, and Yaniv Ziv.
\newblock Time and experience differentially affect distinct aspects of hippocampal representational drift.
\newblock \emph{Neuron}, 111\penalty0 (15):\penalty0 2357--2366.e5, August 2023.
\newblock ISSN 0896-6273.
\newblock \doi{10.1016/j.neuron.2023.05.005}.

\bibitem[Deitch et~al.(2021)Deitch, Rubin, and Ziv]{deitch_representational_2021}
Daniel Deitch, Alon Rubin, and Yaniv Ziv.
\newblock Representational drift in the mouse visual cortex.
\newblock \emph{Current Biology}, 31\penalty0 (19):\penalty0 4327--4339.e6, October 2021.
\newblock ISSN 0960-9822.
\newblock \doi{10.1016/j.cub.2021.07.062}.
\newblock Publisher: Elsevier.

\bibitem[Rule et~al.(2020)Rule, Loback, Raman, Driscoll, Harvey, and O'Leary]{rule_stable_2020}
Michael~E Rule, Adrianna~R Loback, Dhruva~V Raman, Laura~N Driscoll, Christopher~D Harvey, and Timothy O'Leary.
\newblock Stable task information from an unstable neural population.
\newblock \emph{eLife}, 9:\penalty0 e51121, July 2020.
\newblock ISSN 2050-084X.
\newblock \doi{10.7554/eLife.51121}.
\newblock Publisher: eLife Sciences Publications, Ltd.

\bibitem[Markram et~al.(1997)Markram, Lübke, Frotscher, and Sakmann]{markram_regulation_1997}
Henry Markram, Joachim Lübke, Michael Frotscher, and Bert Sakmann.
\newblock Regulation of {Synaptic} {Efficacy} by {Coincidence} of {Postsynaptic} {APs} and {EPSPs}.
\newblock \emph{Science}, 275\penalty0 (5297):\penalty0 213--215, January 1997.
\newblock \doi{10.1126/science.275.5297.213}.
\newblock Publisher: American Association for the Advancement of Science.

\bibitem[Bi and Poo(2001)]{bi_synaptic_2001}
Guo-qiang Bi and Mu-ming Poo.
\newblock Synaptic {Modification} by {Correlated} {Activity}: {Hebb}'s {Postulate} {Revisited}.
\newblock \emph{Annual Review of Neuroscience}, 24\penalty0 (1):\penalty0 139--166, 2001.
\newblock \doi{10.1146/annurev.neuro.24.1.139}.
\newblock \_eprint: https://doi.org/10.1146/annurev.neuro.24.1.139.

\bibitem[Buchanan and Mellor(2010)]{buchanan_activity_2010}
Katherine Buchanan and Jack Mellor.
\newblock The activity requirements for spike timing-dependent plasticity in the hippocampus.
\newblock \emph{Frontiers in Synaptic Neuroscience}, 2, 2010.
\newblock ISSN 1663-3563.

\bibitem[Gilson and Fukai(2011)]{gilson_stability_2011}
Matthieu Gilson and Tomoki Fukai.
\newblock Stability versus {Neuronal} {Specialization} for {STDP}: {Long}-{Tail} {Weight} {Distributions} {Solve} the {Dilemma}.
\newblock \emph{PLOS ONE}, 6\penalty0 (10):\penalty0 e25339, October 2011.
\newblock ISSN 1932-6203.
\newblock \doi{10.1371/journal.pone.0025339}.
\newblock Publisher: Public Library of Science.

\bibitem[Gilson et~al.(2011)Gilson, Masquelier, and Hugues]{gilson_stdp_2011}
Matthieu Gilson, Timothée Masquelier, and Etienne Hugues.
\newblock {STDP} {Allows} {Fast} {Rate}-{Modulated} {Coding} with {Poisson}-{Like} {Spike} {Trains}.
\newblock \emph{PLOS Computational Biology}, 7\penalty0 (10):\penalty0 e1002231, October 2011.
\newblock ISSN 1553-7358.
\newblock \doi{10.1371/journal.pcbi.1002231}.
\newblock Publisher: Public Library of Science.

\bibitem[Carlson et~al.(2013)Carlson, Richert, Dutt, and Krichmar]{carlson_biologically_2013}
Kristofor~D. Carlson, Micah Richert, Nikil Dutt, and Jeffrey~L. Krichmar.
\newblock Biologically plausible models of homeostasis and {STDP}: {Stability} and learning in spiking neural networks.
\newblock In \emph{The 2013 {International} {Joint} {Conference} on {Neural} {Networks} ({IJCNN})}, pages 1--8, August 2013.
\newblock \doi{10.1109/IJCNN.2013.6706961}.
\newblock ISSN: 2161-4407.

\bibitem[Effenberger et~al.(2015)Effenberger, Jost, and Levina]{effenberger_self-organization_2015}
Felix Effenberger, Jürgen Jost, and Anna Levina.
\newblock Self-organization in {Balanced} {State} {Networks} by {STDP} and {Homeostatic} {Plasticity}.
\newblock \emph{PLoS Computational Biology}, 11\penalty0 (9):\penalty0 e1004420, September 2015.
\newblock ISSN 1553-734X.
\newblock \doi{10.1371/journal.pcbi.1004420}.

\bibitem[Caporale and Dan(2008)]{caporale_spike_2008}
Natalia Caporale and Yang Dan.
\newblock Spike {Timing}–{Dependent} {Plasticity}: {A} {Hebbian} {Learning} {Rule}.
\newblock \emph{Annual Review of Neuroscience}, 31\penalty0 (1):\penalty0 25--46, 2008.
\newblock \doi{10.1146/annurev.neuro.31.060407.125639}.
\newblock \_eprint: https://doi.org/10.1146/annurev.neuro.31.060407.125639.

\bibitem[Morrison et~al.(2008)Morrison, Diesmann, and Gerstner]{morrison_phenomenological_2008}
Abigail Morrison, Markus Diesmann, and Wulfram Gerstner.
\newblock Phenomenological models of synaptic plasticity based on spike timing.
\newblock \emph{Biological Cybernetics}, 98\penalty0 (6):\penalty0 459, 2008.
\newblock \doi{10.1007/s00422-008-0233-1}.
\newblock Publisher: Springer.

\bibitem[Bi and Poo(1998)]{bi_synaptic_1998}
Guo-qiang Bi and Mu-ming Poo.
\newblock Synaptic {Modifications} in {Cultured} {Hippocampal} {Neurons}: {Dependence} on {Spike} {Timing}, {Synaptic} {Strength}, and {Postsynaptic} {Cell} {Type}.
\newblock \emph{Journal of Neuroscience}, 18\penalty0 (24):\penalty0 10464--10472, December 1998.
\newblock ISSN 0270-6474, 1529-2401.
\newblock \doi{10.1523/JNEUROSCI.18-24-10464.1998}.
\newblock Publisher: Society for Neuroscience Section: ARTICLE.

\bibitem[Haynes(2013)]{haynes_wilcoxon_2013}
Winston Haynes.
\newblock Wilcoxon {Rank} {Sum} {Test}.
\newblock In Werner Dubitzky, Olaf Wolkenhauer, Kwang-Hyun Cho, and Hiroki Yokota, editors, \emph{Encyclopedia of {Systems} {Biology}}, pages 2354--2355. Springer, New York, NY, 2013.
\newblock ISBN 978-1-4419-9863-7.
\newblock \doi{10.1007/978-1-4419-9863-7_1185}.

\bibitem[Cohen and Kohn(2011)]{cohen_measuring_2011}
Marlene~R. Cohen and Adam Kohn.
\newblock Measuring and interpreting neuronal correlations.
\newblock \emph{Nature Neuroscience}, 14\penalty0 (7):\penalty0 811--819, July 2011.
\newblock ISSN 1546-1726.
\newblock \doi{10.1038/nn.2842}.
\newblock Number: 7 Publisher: Nature Publishing Group.

\end{thebibliography}

\onecolumn
\newpage

%%%%%%%%%%%%%%%%%%%%%%%%%%%%%
% Supplementary Information %
%%%%%%%%%%%%%%%%%%%%%%%%%%%%%
\captionsetup*{format=largeformat}

\setcounter{figure}{0}
\renewcommand{\thefigure}{S.\arabic{figure}}

\section{Relation between the GMR process and Loewenstein et al.'s model} \label{app:1}

Let us begin by introducing the phenomenological model presented by Loewenstein et al. in \cite{loewenstein_multiplicative_2011}, which describes the dynamics of the logarithm of the size of the $k$-th dendritic spine in a given synapsis, $\tilde{X}^k$, as a sum of two independent Ornstein-Uhlenbeck processes: 
\begin{equation}
    \log_{10}(\tilde{X}^k) = Y_1^{k} + Y_2^{k} + \log_{10}\tilde{ \mu} \ .
\end{equation}
In the above equation, $\tilde{ \mu}$ is the average of all spine sizes and $Y_1^{k}$ and $Y_2^{k}$ are OU processes described by:
\begin{equation}
\tau_i \frac{dY_i^k}{dt} = -Y_i^l + \sqrt{2 \tau_{i} 
}\tilde{ \sigma}_{i} \xi_i^k,  \quad i=1,2 \label{eq:app_OU_Process}
\end{equation} where $\xi_i^k$ is a Gaussian white noise such that   $\langle \xi_i^k \rangle=0$ and $\langle \xi_i^k(t) \xi_j^{k^{\prime}}(t') \rangle = \delta_{i,j}\delta_{k,k^\prime} \delta(t-t')$. In the above model, the values for the timescales of the two processes, $\tau_1$ and $\tau_2$, as well as their stationary variance, $\tilde{\sigma}_1$ and $\tilde{\sigma}_2$, where fitted using empirical measures of the spine sizes in the dendrites of auditory cortex neurons recorded \emph{in vivo} in mice. For our purposes, let us further simplify the above model considering that the dynamics is characterized by a single, most-relevant timescale, such that:
\begin{equation}
    \ln(\tilde{X}^k) = Y^{k} + \ln\tilde{ \mu} \ , \label{eq:app_eq3}
\end{equation}
where, without loss of generality, we changed the decimal logarithm for the natural one to simplify the upcoming analysis. 

Let us now go back to the stochastic differential equation for a geometric mean-reverting process:
\begin{equation}
\dfrac{dX^k}{dt} = \omega (\mu - X^k) + \sigma X^k \xi^k \ .
\end{equation}
Applying Ito's Lemma to the function $\ln(X^k)$:
\begin{equation}
   \dfrac{d(\ln(X^k))}{dt}  = \omega \left( \frac{\mu}{X^k} - 1 \right) - \frac{1}{2} \sigma^2 + \sigma \xi^k. \label{eq:app_eq5}
\end{equation}
In order to check the relation between the dynamics of both processes, let us assume that synaptic weights in our model, $X^k$, describe a magnitude proportional to the spine sizes empirically measured in \cite{loewenstein_multiplicative_2011} (i.e., $X^k\equiv \alpha\tilde{X}^k$). Therefore, using Eq. \ref{eq:app_eq3} we can approximate:
\begin{equation}
    \frac{\mu}{X^k} \approx \frac{1 - Y^k}{\alpha} \ , 
\end{equation}
which, substituting in Eq.~~\ref{eq:app_eq5}, leads to:
\begin{equation}
    \frac{dY}{dt} = \omega \left( \frac{\mu}{\alpha\tilde{\mu}}\left( 1 - Y^k \right) - 1 \right) - \frac{1}{2} \sigma^2  + \sigma \xi^k \ .
\end{equation}
Comparing term by term with Eq.~\ref{eq:app_OU_Process} for $i=1$, and after several manipulations,  one can finally see that both, the GMR process and the phenomenological model proposed in \cite{loewenstein_multiplicative_2011} (although reduced to a single timescale), follow identical equations under the following re-scaling of parameters:
\begin{align}
   \sigma &= \sqrt{ \frac{2}{\tau}}\tilde{\sigma} \\
   \omega  &= \frac{1 - \tilde{\sigma}^2}{\tau}  \\
    \mu &= \frac{\tilde{\mu}\alpha}{1 - \tilde{\sigma}^2}  \ .
\end{align}

\newpage

\section{The effects of learning-induced deterministic change in representation: A simple model} \label{toymodel}

As we showed in the main text, changes in the neural representation of a particular odor occur through two mechanisms: learning induces synaptic changes towards the odor-specific sub-manifold in a deterministic fashion, whereas the stochastic weight dynamics causes random fluctuations akin to a Brownian motion.  Here, we develop a simple model to demonstrate the interplay of these two mechanisms and show how they affect the measured representational ``drift''. 

Let us consider a two-dimensional representational space $(x,y)$ such that,  for a particular odor, the representational sub-manifold is a 1-D line at $x=0$. Starting from an initial response $(x_0,y_0)$ at time $t=0$, the odor is presented at time $t_i=(i-1)\tau$ $(i=1,2,...)$, where $\tau$  is the time interval between presentations.  During the short presentation of the odor, the effect of learning drives the representation towards the representation sub-manifold, i.e., the $x=0$ line. Apart from these brief odor presentations, the representation changes randomly. To capture these two effects, we can formulate a simple model to describe the dynamics of the representation:
\begin{eqnarray}
    \frac{dx}{dt}&=&f(x)\sum_{i=1}^n\delta(t-t_i)+\eta_x(t),\label{dxdt}\\
    \frac{dy}{dt}&=&\eta_y(t),\label{dydt}
\end{eqnarray}
where the first term in the right-hand side of Eq.~\ref{dxdt} represents the effects of learning during the brief odor presentation with $f(x=0)=0$ and $f'(x=0)<0$, which guarantees that $x=0$ is the fixed line for the learning dynamics; while  $\eta_{x}(t)$ and  $\eta_{y}(t)$ are random white noises mimicking the effect of stochastic weight changes: $\langle \eta_{x,y}\rangle=0$, $\langle \eta_x (t')\eta_x(t)\rangle=2\Omega_x\delta(t-t')$, $\langle \eta_y (t')\eta_y(t)\rangle=2\Omega_y\delta(t-t')$. For simplicity, we set $f(x)=-\alpha x$ where $\alpha>0$ measures  the strength of learning. We also assume $\Omega_x=\Omega_y=\Omega$, which is the strength of the random drift. The measured representational drift after $n$ odor presentations or total time $T=(n-1)\tau$ is given by: $\sigma^2=\sigma_x^2+\sigma_y^2$ where  $\sigma_x^2\equiv \langle (x_0-x(T))^2\rangle$ and $\sigma_y^2\equiv \langle (y_0-y(T))^2\rangle$. 

In the direction normal to the representational sub-manifold (i.e., the $y$-direction) learning does not play any role and we have $\sigma_y^2=\Omega T$. On the other hand, learning does play an important role in the $x$-direction and $\sigma_x^2$ can be computed as follows. Denote the values of $x$ before and after the odor presentation at time $t_i$ as $x_i^-$ and $x_i^+$, respectively. By integrating  Eq.~\ref{dxdt} from $t_i^-=t_i-\epsilon$ to $t_i^+=t_i+\epsilon$ in an infinitesimal region around time $t_i$ of the odor presentation, we have: $\int_{t_i^-}^{t_i^+}\frac{dx}{x}=\ln(x_i^+/x_i^-)=-\alpha$, which leads to the equation for the ``jump" process:
\begin{equation}
    x_i^+=e^{-\alpha}x_i^-.
    \end{equation}

Between two consecutive odor presentations, the dynamics is a pure diffusion process, and we have:
\begin{equation}
       x_{i+1}^-=x_i^++\xi_i \ ,
\end{equation}
where $\xi_i=\int_{t_i}^{t_{i+1}}\eta_x(t)dt$ is the displacement in $x$ due to random weight change during time $t_i$ to $t_{i+1}$. It is easy to show that $\langle\xi_i\rangle=0$ and $\langle\xi_i^2\rangle=\Omega \tau$.  Together with the initial condition, $x_1^-=x_0$, we can the write:
\begin{equation}
x(T)\equiv x_n^+=  e^{-n\alpha}x_0+\sum_{i=2}^n e^{-(n-i+1)\alpha}\xi_{i-1} \ ,
\end{equation}
from which we obtain:
\begin{equation}
    \sigma_x^2\equiv \langle (x_0-x(T))^2\rangle=x_0^2(1-e^{-n\alpha})^2+\Omega \tau \frac{e^{-2\alpha}-e^{-2n\alpha}}{1-e^{-2\alpha}} \ .
    \label{sigma_x}
\end{equation}

In the above expression, the first term comes from the deterministic changes of the neural activity from the initial position $x_0$ towards the sub-manifold, $x=0$; whereas the second term comes from the random drift damped by learning. The behavior of $\sigma_x^2$ in the absence or presence of learning is summarized below:  

\begin{itemize}
    \item In the absence of learning (i.e.,  $\alpha\rightarrow 0$), from Eq.~\ref{sigma_x} we have: $\sigma_x^2=\Omega \tau (n-1)=\Omega T$, which is just random diffusion as $\sigma_x^2$ is linearly proportional to time.
    
    \item In the presence of learning (i.e., for a finite $\alpha$) the measured ``drift'', $\sigma_x^2$, is suppressed. Specifically, although $\sigma_x^2$ increases with $n$ transiently, it saturates to a constant when $n\alpha\gg 1$: $\sigma_x^2(n\gg \alpha^{-1})=x_0^2+\Omega \tau(1-e^{-2\alpha})^{-1}$. %{\bf This is what happens in the piriform cortex as we explained in our paper. }
\end{itemize}   

Overall, even though learning suppresses drift at long time (or large values of $n$), there may be an intermediate range of $n$ where the drift measured by $\sigma_x^2$ is actually bigger with a finite $\alpha$. The reason is that even though a finite $\alpha$ always suppresses the second term in Eq.~\ref{sigma_x}, it actually enhances the first term that is proportional to $x_0^2$ (notice that the first term vanishes in the absence of learning). Therefore, depending on the initial distance to the representational sub-manifold, $x_0$, the deterministic changes towards the line $x=0$ can dominate the measured ``drift'', $\sigma_x^2$, effectively making it larger than that without learning. This is what we refer to in the Discussion section of the main text as \emph{fast drift} \cite{khatib_active_2023}.

\section{Alignment of fast stimulus-induced drift with the directions of noise variability}\label{NoiseAlignment}

As we showed throughout this work, the actual slow random drift, which we hypothesize stems from noisy multiplicative synaptic dynamics, affects the tuning of single neurons to a given set of stimuli (Fig.\ref{fig:2}i-j), thus affecting the measured \emph{signal correlation} across days (i.e., pair-wise correlations between mean responses to different stimuli, see Fig.\ref{fig:SI_DriftMeasures}k). 

In contrast, \emph{noise correlations} measure the observed co-variability in neural responses to the same stimulus repeated across different trials, under the same behavioral conditions \cite{cohen_measuring_2011}. From an experimental point of view, the effects of STDP-induced changes ---or, as we called it, the fast stimulus-dependent drift--- , which can take place on a scale of milliseconds to seconds between trials, would emerge in long-recording experiments as part of this \emph{noise variance} or trial-to-trial variability in the responses to the same stimulus set within a given test day.  

\begin{figure*}
\centering{}
\includegraphics[width=0.9\linewidth]{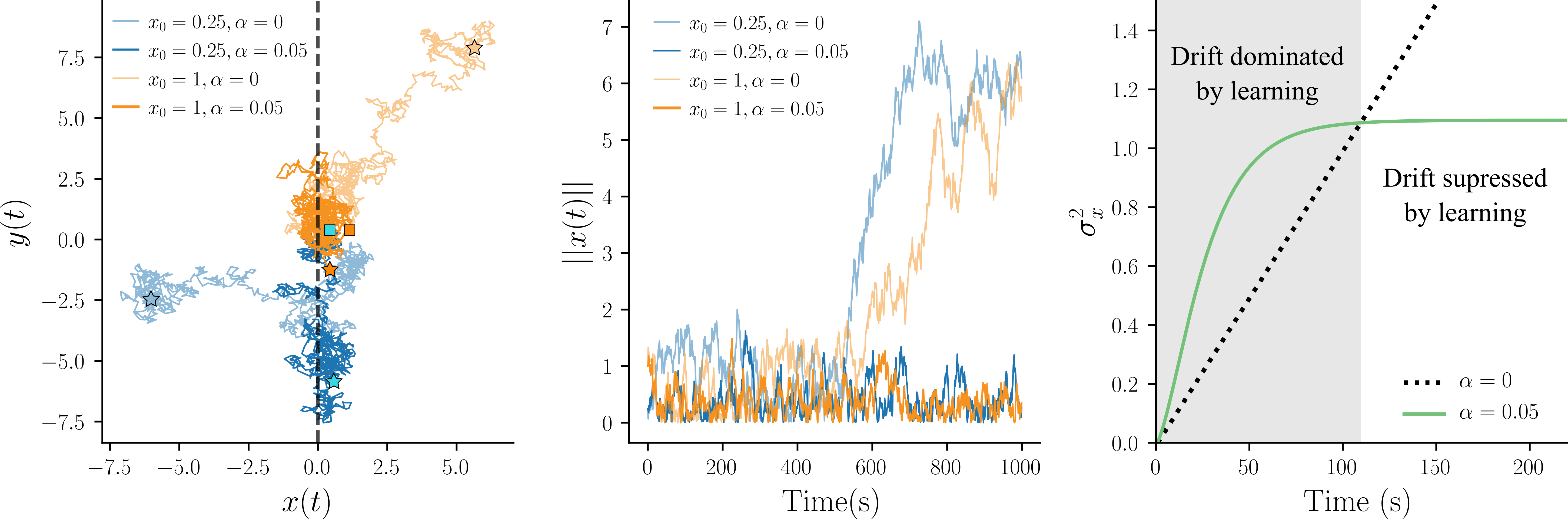}
\caption{\textbf{Drift alignment with ``noise'' variability} \textbf{(a)} Drift (black arrow) and noise (green arrow) directions in a 2-dimensional subspace spanned by population responses to two odors: each circle represents population mean activity on a single trial. For each pair of test days, given an odor, $o$, and trial, $m$, a drift vector $\Delta \vec{\mu}_{o,m}$ can be defined between the trial-conditioned population vectors. \textbf{(b)} Drift alignment with noise direction averaged over all unfamiliar odors, presented every 8 days (red); over all familiar odors, presented every day (blue); and the same quantity for the empirical measure in Rule \emph{et al.} \cite{rule_stable_2020}. Dotted black line shows the expected chance alignment for two random vectors in an $N$-dimensional space.}\label{fig:SI_AngleOverlap}
\end{figure*}

Thus, if our hypothesis is right and the overall observed drift can have a contribution from fast, learning-induced dynamics at short time scales, then for every input and pair of recorded days, a general \emph{drift vector} (pointing in the direction of population changes between such days in the N-dimensional space) should still have a small but significant projection in the directions encoding noise variability (see Fig. \ref{fig:SI_AngleOverlap}a).  Remarkably, this is exactly what Rule et al. observed when analyzing long-term calcium imaging recordings from mice in the mouse posterior parietal cortex (PPC) during a virtual reality T-maze task \cite{rule_stable_2020} (Fig.\ref{fig:SI_AngleOverlap}b, gray box). 

To test whether whether the same phenomenon can be captured with our model, we measured the alignment of the drift with the direction of maximum trial-to-trial variability by adapting the protocol introduced in \cite{rule_stable_2020}. Let us define a \emph{drift vector}, $\Delta \vec{\mu}_{o,m}^{pq}=\mathbf{x}_{o,m, p}-\mathbf{x}_{o,m, q}  $, as the trial-conditioned change in the mean firing rate response of the population to odor $o$, during trial $m$, between days $p$ and $q$. We now compute, for each odor, $o$, the noise or trial-to-trial covariance matrix on the first test day, $\Sigma_o$, and define an statistic that determines the amount of overlapping between the drift vector and the noise encoding direction as:
\begin{equation}
    {\phi_{o,m}^{pq}}^2 = \dfrac{{\Delta \vec{\mu}_{o,m}^{pq}}^\intercal \Sigma_o \Delta \vec{\mu}_{o,m}^{pq}}{\lambda_{max}\lvert \Delta \vec{\mu}_{o,m}^{pq}  \rvert ^2} \ , 
\end{equation}
where we normalized the drift vector to unit length and divided by the maximum eigenvalue, $\lambda_{max}$, of the noise covariance matrix. Thus, the above statistic provides a measure of the amount of drift that can be explained (or overlaps) with the principal direction of noise correlations, being one if both vectors align perfectly, and zero if they are completely orthogonal \cite{rule_stable_2020}. Finally, to discount for the expected chance alignment that can take place between any two random vectors in an $N$-dimensional space, $\psi_0 = \langle \mathrm{tr}(\Sigma_o)/(N\lambda_{max}) \rangle$ (where the average is taken over all odors, see also \cite{rule_stable_2020} for more details), we compute a corrected final estimate as: 
\begin{equation}
    \rho_{o,m}^{pq} = \dfrac{\phi_{o,m}^{pq} - \psi_0 }{1-\phi_{o,m}^{pq}} \ .
\end{equation}

Figure \ref{fig:SI_AngleOverlap}b shows the distribution of the above measure across all trials and odors, for the case in which we have unfamiliar odors presented every 8 days (red boxplot) and familiar odors presented every day (blue boxplot). As we can see, non trivial overlapping is observed for representations of familiar odors, in which the presentation of the stimulus ``pulls back" the weights towards the submanifold previously learned during the familiarization process. The same quantity for the PPC, as measured in \cite{rule_stable_2020}, is shown for comparison purposes (gray boxplot).

%On the other hand, the  reinforcement of selective connections during stimulus presentation, mediated by STDP effects,  would indeed explain the observed sparsification of the neural code on a shorter timescale both in experiments \cite{khatib_active_2023} and simulations \cite{ratzon_representational_2024}. 

%Moreover, just as it was shown in \cite{rule_stable_2020}, we also observe the existence of a smaller but still significant overlapping of the drift vector with the direction of signal correlations (i.e., directions of neural activity encoding information about the represented stimuli). We hypothesize that this overlap is actually a result of the structural constraints in the network (i.e., connections are sparse and weight changes are only allowed in existing synapses), which a null-model based in the expected correlations between random N-dimensional vectors would not capture. 

\newpage

\section*{Supplementary Figures} 

\begin{figure*}[htbp]
\centering{}
\includegraphics[width=0.75\linewidth]{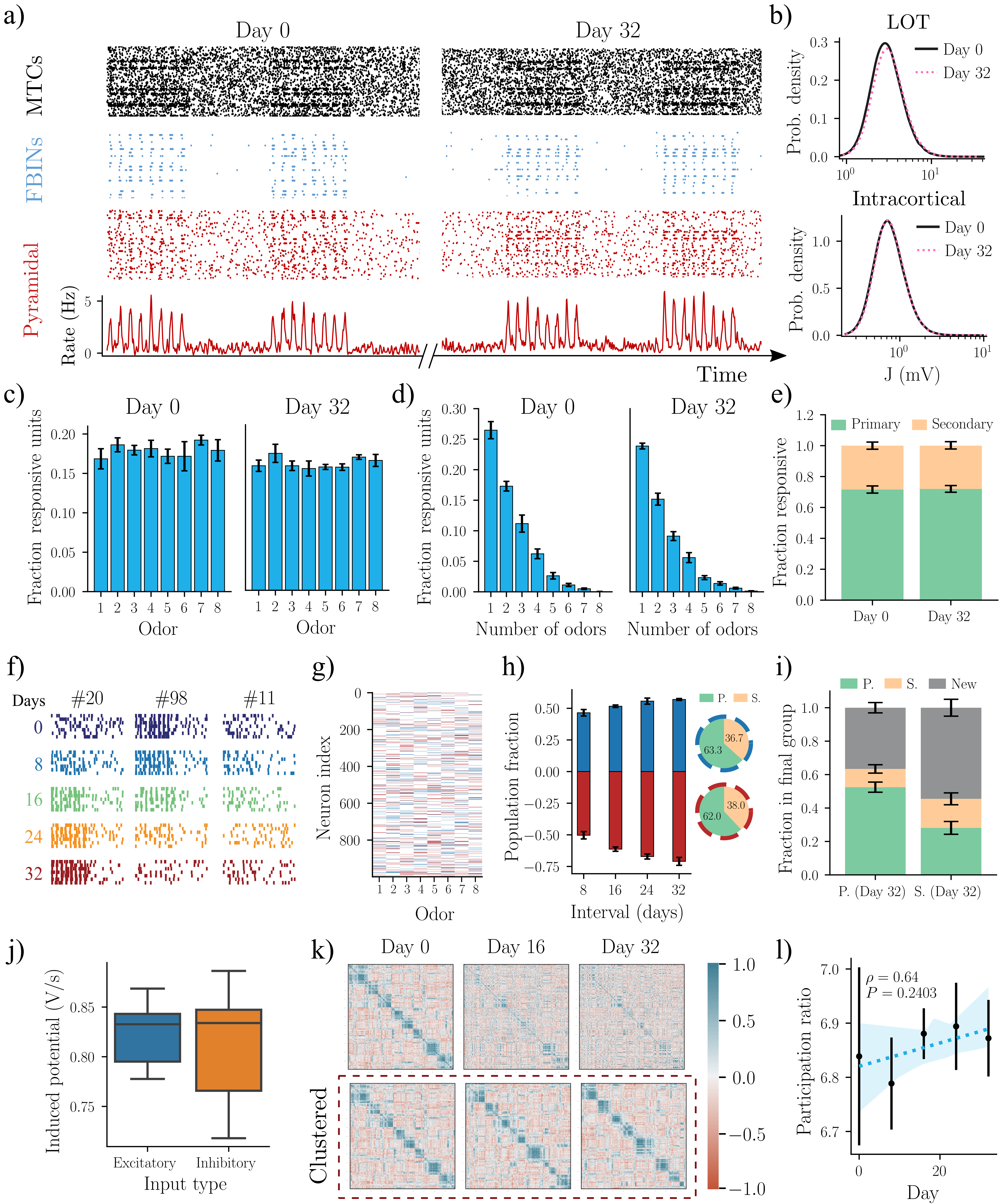}
\caption{\textbf{Extended results on the effects of representational drift.} \textbf{(a)} Raster plots for a sub-sample ($10\%$) of each neural population during two trials of an odor presentation on day $0$ and day $32$, together with the corresponding pyramidal population average firing rate. \textbf{(b)} Lognormal probability distributions for MTC-to-pyramidal (top) and pyramidal-to-pyramidal(bottom) weights. \textbf{(c)} Fraction of pyramidal neurons that respond to each odor. \textbf{(d)} Fraction of pyramidal neurons responsive to \emph{exactly} $n$  odors. \textbf{(e)} Fraction of primary and secondary labels within responsive units. \textbf{(f)} Raster plots for the same three odor-unit pairs in Fig.~\ref{fig:2}j across test days. Each rows represents a single trial of odor presentation. \textbf{(g)} Changes in neuron selectivity after $32$ days (blue=gain; red=loss) to each odor with respect to the beginning of the experiment. \textbf{(h)} Left: fraction of units that gained (blue) or lost (red) responsiveness, relative to the total number of initially responsive units. Right, top: fraction of primary and secondary labels for all units responsive on last but not first test day. Right, bottom: same as previous, but for all units responsive on first but not last day. \textbf{(i)} Composition of primary and secondary sets on day $32$ in terms of unit classes on day $0$. \textbf{(j)} Change to pyramidal membrane potential from excitatory (blue) and inhibitory (orange) sources over $1\si{s}$, illustrating the existing balance between excitation and inhibition in our model. \textbf{(k)} Pearson's correlation matrices for pyramidal responses on three test days, with units clustered by correlation on day $0$ (top), or correlation in each test day (bottom). \textbf{(l)} Dimensionality (as measured by the normalized participation ratio) of drifting representation manifold, averaged over $n=6$ experiments (mean $\pm$ std, shaded blue area: 95\% CI for regression).}\label{fig:SI_DriftMeasures}
\end{figure*}

\begin{figure*}
\centering{}
\includegraphics[width=0.95\linewidth]{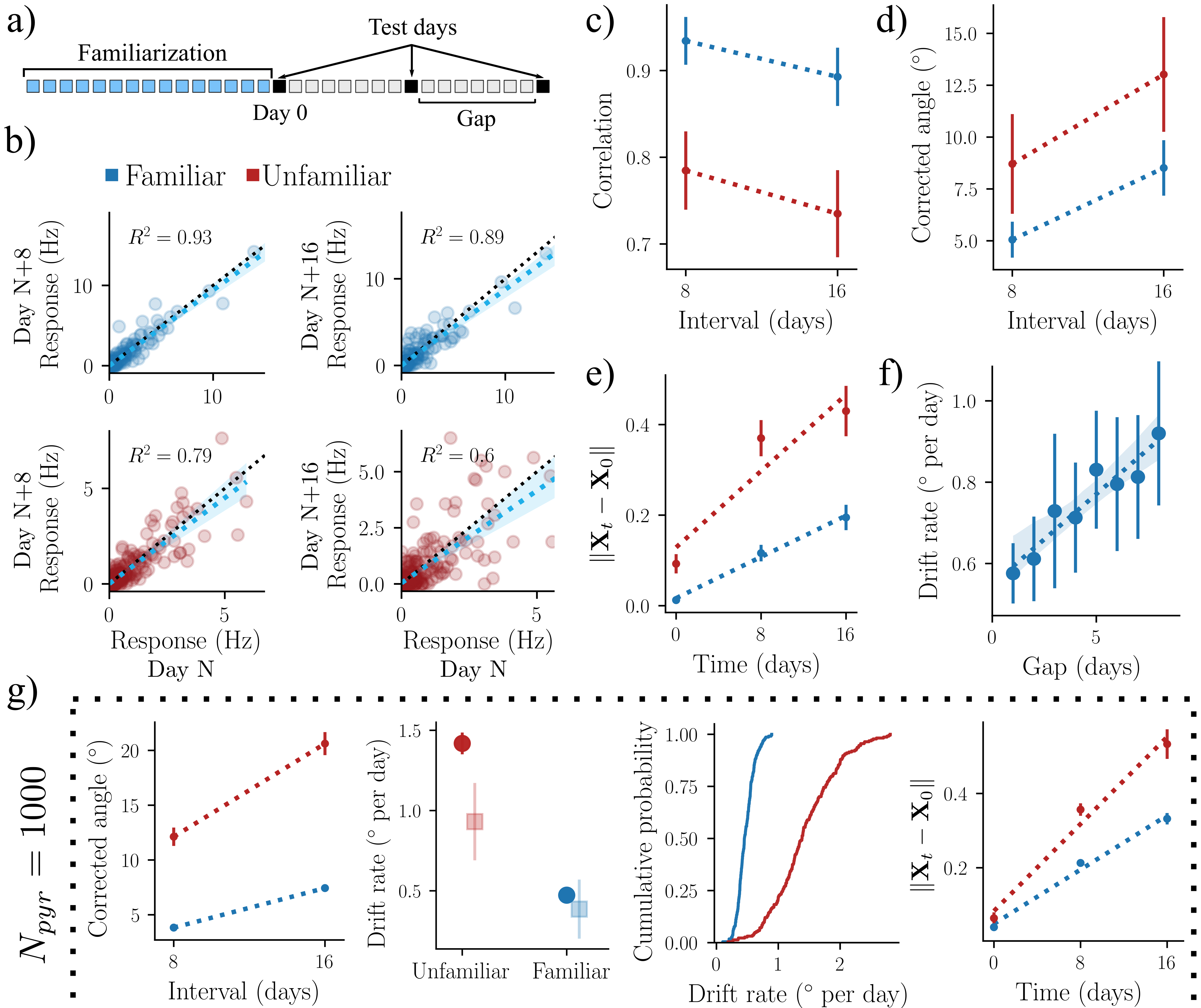}
\caption{\textbf{Extended results on the effects of learning.} In all panels, unless otherwise stated, simulations mimic experiments with Cohort A, with blue (red) denoting familiar (unfamilar) odors. \textbf{(a)} Schematic depiction of a typical simulated experiment. \textbf{(b)} Regression of of firing rate responses across $8$- to $32$-days intervals for 100 randomly chosen odor-unit pairs. \textbf{(c)} Average population vector correlation and \textbf{(d)} average corrected angle between same-odor representations across intervals between test days.  \textbf{(e)} Euclidean distance between same-odor representations across test days, averaged over odors and trials, and normalized by average within-day distance for different odor representations. \textbf{(f)} Average drift rate in degree angles per day for simulations interpolating between Cohort A (odors presented every day, first point) and Cohort B (odors presented every 8 days, last point). \textbf{(g)} From left to right: corrected angle, average drift rate, cumulative probability distribution and Euclidean distance between same-odor representations for a network with $N=1000$ pyramidal neurons. 
}\label{fig:SI_FamVSUnf}
\end{figure*}

\begin{figure*}
\centering{}
\includegraphics[width=0.95\linewidth]{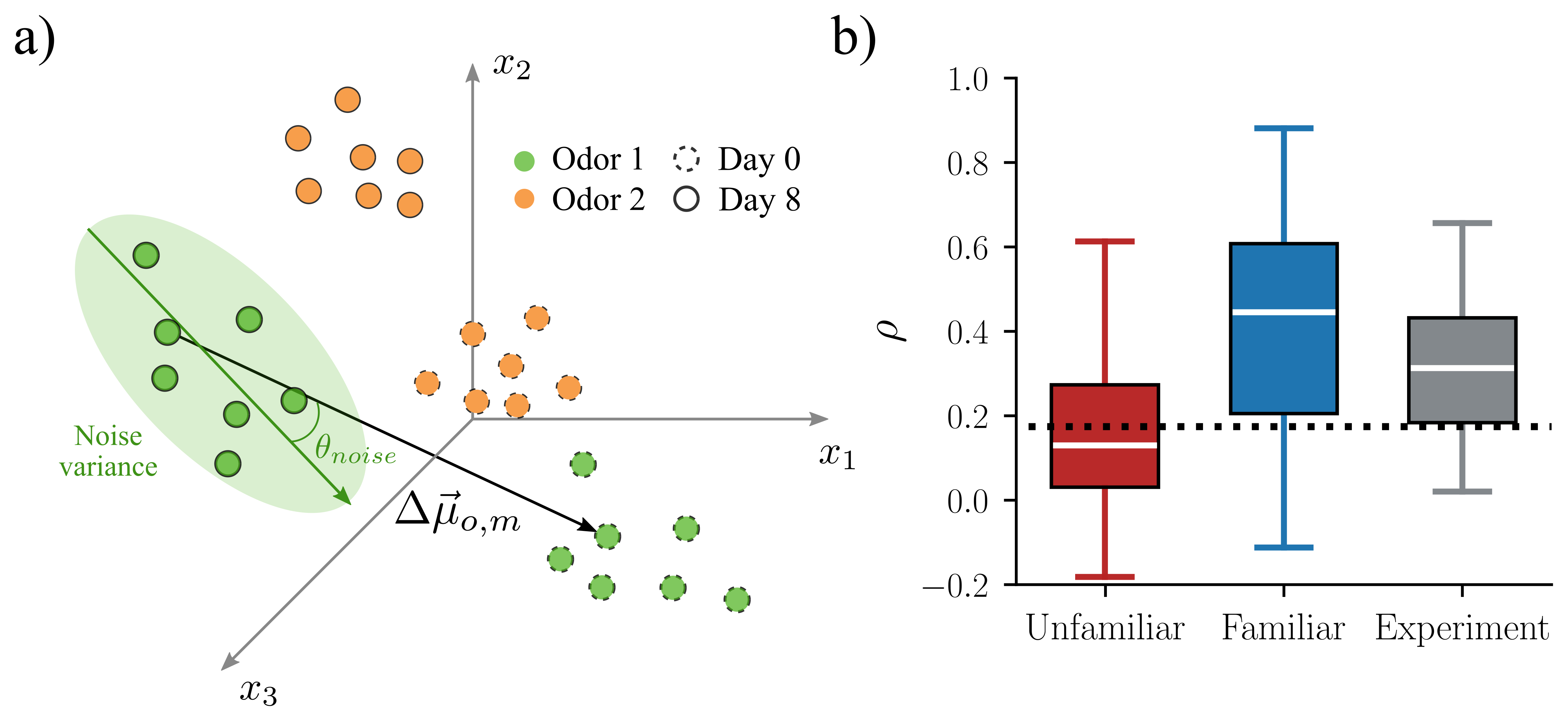}
\caption{\textbf{A simple model for the interplay between learning and drift.}  \textbf{(a)} Trajectories in the $(x, y)$ representational space for the model described by Eqs.\ref{dxdt}-\ref{dydt} with a ``learned'' submanifold at $x=0$ (green dashed line) and two different initial conditions: $x_0 = 0.25$ (blue) and $x_0 = 1$ (orange).  For each initial configuration (square marker), we ran simulations with ($\alpha=0.05$) and without ($\alpha=0$) learning.  Star markers point to the end of the trajectories.  \textbf{(b)} As in  (a), but now we plot the distance to the $x=0$ sub-manifold against the simulated time.  \textbf{(c)} Solution to Eq.\ref{sigma_x} for an initial condition $x_0=1$, with learning ($\alpha=0.05$, green line) and without learning ($\alpha=0$, black dotted line). In all simulations: $T=1000\si{s}$, $\tau=1\si{s}$, $\Omega=0.01$, $y_0=0.5$. For plots in (a) and (b), trajectories are integrated with a time step $dt=0.002$ and sampled every $500$ steps.}\label{fig:SI_SimpleModel}
\end{figure*}

\end{document}